\documentclass[11pt,a4paper]{article}
\usepackage{jcappubnew}
\usepackage{cancel}

\newcommand{\beq}{\begin{equation}}
\newcommand{\eeq}{\end{equation}}
\newcommand{\bea}{\begin{eqnarray}}
\newcommand{\eea}{\end{eqnarray}}

\newcommand{\vc}[1]{{\boldsymbol #1}}

\newcommand{\lh}{\left(}
\newcommand{\rh}{\right)}
\newcommand{\der}{\partial}
\renewcommand{\d}{\mathrm{d}}

\DeclareMathSymbol{\mg}{\mathrel}{symbols}{"1D}

\newcommand{\cA}{{\mathcal A}}
\newcommand{\cB}{{\mathcal B}}

\newcommand{\cD}{{\mathcal D}}

\newcommand{\cG}{{\mathcal G}}

\newcommand{\cK}{{\mathcal K}}
\newcommand{\cL}{{\mathcal L}}

\newcommand{\cO}{{\mathcal O}}

\newcommand{\cQ}{{\mathcal{Q}}}

\newcommand{\f}{f_\mathrm{NL}}

\newcommand{\bg}{{\bar g}}
\newcommand{\bh}{{\bar h}}

\newcommand{\bA}{{\bar A}}

\newcommand{\bE}{{\bar E}}

\newcommand{\bG}{{\bar G}}

\newcommand{\bK}{{\bar K}}

\newcommand{\bN}{{\bar N}}

\newcommand{\bP}{{\bar P}}

\newcommand{\bR}{{\bar R}}
\newcommand{\bS}{{\bar S}}

\newcommand{\bX}{{\bar X}}

\newcommand{\bdm}{\begin{displaymath}}
\newcommand{\edm}{\end{displaymath}}
\newcommand{\nn}{\nonumber}
\def\be{\begin{equation}}
\def\ee{\end{equation}}

\def\e{\epsilon}

\def\hpa{\eta^{\parallel}}
\def\thpa{\tilde{\eta}^{\parallel}}
\def\hpe{\eta^{\perp}}
\def\thpe{\tilde{\eta}^{\perp}}
\def\hpas{\eta^{\parallel}_*}

\def\dphi{\dot{\phi}}

\title{Covariant second-order perturbations in generalized two-field inflation}

\author{Eleftheria Tzavara$^{1}$,}
\author{Shuntaro Mizuno$^{2,1}$,}
\author{and Bartjan van Tent$^{1}$}
\affiliation{$^1$Laboratoire de Physique Th\'eorique, Universit\'e Paris-Sud 11 
and CNRS, B\^atiment 210, 91405 Orsay Cedex, France,\\
$^2$Laboratoire Astroparticule et Cosmologie (APC), UMR 7164-CNRS, Universit\'{e} Denis Diderot-Paris 7,
        10 rue Alice Domon et L\'{e}onie Duquet, 75205 Paris, France}
\emailAdd{Eleftheria.Tzavara@th.u-psud.fr} 
\emailAdd{Shuntaro.Mizuno@apc.univ-paris7.fr}
\emailAdd{Bartjan.Van-Tent@th.u-psud.fr}

\begin{document}

\begin{flushright}
LPT-Orsay-13-139\\ 
\end{flushright}

\abstract{We examine the covariant properties of generalized models of
  two-field inflation, with non-canonical kinetic terms and a possibly
  non-trivial field metric. We demonstrate that kinetic-term
  derivatives and covariant field derivatives do commute in a proper
  covariant framework, which was not realized before in the
  literature. We also define a set of generalized slow-roll
  parameters, using a unified notation. Within this framework, we
  study the most general class of models that allows for well-defined 
  adiabatic and entropic
  sound speeds, which we identify as the models with parallel momentum
  and field velocity vectors. For these models we write the exact
  cubic action in terms of the adiabatic and isocurvature
  perturbations. We thus provide the tool to calculate the exact
  non-Gaussianity beyond slow-roll and at any scale for these
  generalized models.  We illustrate our general results by considering
  their long-wavelength limit, as well as with the example of
  two-field DBI inflation.}

\maketitle

\flushbottom

\section{Introduction}

Inflation is widely believed to be the most likely explanation for
the origin of temperature fluctuations of the Cosmic Microwave Background (CMB)
\cite{paper1,paper2,paper3,paper4} 
(see e.g. \cite{Mukhanov:1990me, Lyth:1998xn} for reviews).
The prediction that the primordial density perturbations
generated during inflation are nearly scale-invariant and almost Gaussian
is strongly supported by recent Planck observations
\cite{Ade:2013zuv,Ade:2013uln}. However, due to the limited information 
available in the power spectrum of the primordial density perturbations, 
it is still difficult to determine the most consistent inflation model
among a very large number of models.

Therefore, in order to distinguish otherwise degenerate models,
additional observables are necessary.
In particular, the non-Gaussianity of the primordial density perturbations
has become an increasingly popular observable \cite{Komatsu:2009kd}.
It has been shown to be suppressed by a slow-roll parameter 
in the simplest single-field inflation models \cite{Maldacena:2002vr}.
Hence any observation of primordial non-Gaussianity would be an indication
of a more complicated model.
Although it was reported recently by Planck \cite{Ade:2013ydc} that no
primordial non-Gaussianity of the standard local, equilateral, and orthogonal
shapes has been detected, the higher precision of the Planck constraints means
that it is now more important than ever that each inflation model should pass
this test. Hence the development of methods to compute non-Gaussianity and
a better understanding of the correspondence between theoretical inflation 
models and the type and magnitude of the non-Gaussianity of the CMB remain
very important (see \cite{Bartolo:2004if,Chen:2010xka} for reviews).

There are a number of distinct mechanisms to produce non-Gaussianity
during inflation. One of these occurs in multiple-field inflation where
not only adiabatic perturbations but also entropy perturbations
are produced. It is well known that since the entropy perturbations
can be converted into adiabatic ones 
during\footnote{The non-linearities can also develop through the conversion
of the entropy perturbations into adiabatic perturbations after inflation
as in the curvaton mechanism \cite{Moroi:2001ct,Lyth:2001nq} 
and modulated reheating \cite{Kofman:2003nx,Dvali:2003em}.} 
multiple-field inflation,
the gauge-invariant curvature perturbation
$\zeta$ originally defined in \cite{Bardeen:1980kt} can evolve
on super-horizon scales 
\cite{Starobinsky:1994mh,Gordon:2000hv,GrootNibbelink:2001qt}.
This conversion is associated with a turn of the inflaton trajectory
and can produce non-Gaussianity of the so-called local type.

On the other hand, if the inflaton possesses non-canonical kinetic terms
like k-inflation \cite{ArmendarizPicon:1999rj,Garriga:1999vw} or 
DBI inflation \cite{Silverstein:2003hf,Alishahiha:2004eh},
another type of non-Gaussianity is produced during horizon crossing
by the derivative interactions of the field, even in single-field models
\cite{Chen:2006nt}. This non-Gaussianity is of the so-called equilateral
type \cite{Creminelli:2005hu}.\footnote{Sometimes so-called orthogonal-type 
\cite{Senatore:2009gt} non-Gaussianity is produced instead or in addition.}
Studying multiple-field models with non-canonical terms is therefore very
interesting: one can in principle produce both local and equilateral 
non-Gaussianity \cite{RenauxPetel:2009sj}.

The first study of the third-order action in multiple-field inflation
was done in \cite{Seery:2005gb} for models with canonical
kinetic terms. However, since the action of \cite{Seery:2005gb} is written
in terms of the scalar fields, the authors had to use the $\delta N$ formalism
\cite{Starobinsky:1986fxa,Sasaki:1995aw,Sasaki:1998ug,Wands:2000dp,Lyth:2004gb,Lyth:2005fi}
to compute the non-Gaussianity of the adiabatic perturbation $\zeta$ that 
is directly related to the non-Gaussianity of the CMB.
The standard $\delta N$ formalism requires that the slow-roll approximation is imposed 
at horizon crossing in order to be able to neglect spatial gradients outside 
the horizon  \cite{Leach:2001zf} (for generalizations of the $\delta N$ 
formalism see \cite{Takamizu:2010xy,Lee:2005bb}). 
In \cite{Tzavara:2011hn}, two of us succeeded in constructing
the exact third-order action directly for the gauge-invariant second-order 
adiabatic and isocurvature perturbations (instead of the fields), which is valid
beyond any slow-roll or super-horizon approximations. This included deriving
a proper definition of the second-order gauge-invariant isocurvature 
perturbation and resolving several issues regarding gauge invariance at 
second order. (For discussions on the gauge-invariant curvature 
perturbation at second order, see also
\cite{Bruni:1996im,Noh:2003yg,Rigopoulos:2003ak,Malik:2003mv,Lyth:2005du,Malik:2005cy,Langlois:2005qp,Nakamura:2004rm}.)
The analysis of \cite{Tzavara:2011hn} is based on the  
the covariant\footnote{Strictly speaking, 
it is the formalism proposed by \cite{Langlois:2005qp} that
is called the ``covariant formalism".
But since the formalism developed by RSvT
is based on the covariant approach \cite{Ellis:1989jt}
and it is in principle
equivalent to the covariant formalism after choosing a coordinate system
and expanding to the given order in spatial gradients, 
we simply refer to the formalism developed by
RSvT as ``covariant approach" here.} 
approach
\cite{Rigopoulos:2004ba,Rigopoulos:2005xx,Rigopoulos:2005ae,Rigopoulos:2005us,Tzavara:2010ge}
developed by Rigopoulos, Shellard and Van Tent, hereafter referred to as RSvT.
(For related works on non-Gaussianity in multiple-field inflation models
with canonical kinetic terms, see
\cite{Bernardeau:2002jy,Vernizzi:2006ve,Kim:2006te,Battefeld:2006sz,Choi:2007su,Battefeld:2007en,Yokoyama:2007uu,Yokoyama:2007dw,Sasaki:2008uc,Cogollo:2008bi,Naruko:2008sq,Byrnes:2008wi,Langlois:2008vk,Byrnes:2009qy,Battefeld:2009ym,Byrnes:2009pe,Byrnes:2010ft,Wang:2010si,Meyers:2010rg,Peterson:2010mv,Elliston:2011dr,Watanabe:2011sm,Frazer:2011br,Choi:2012hea,Mazumdar:2012jj,Tzavara:2012qq}.)

In this paper we extend the work of \cite{Tzavara:2011hn} to the case of
two-field models with general kinetic terms and a possibly non-trivial field
metric.
This is motivated from the point of view of inflation models
based on high-energy theories like supergravity or string theory, where 
the inflaton fields 
are identified as K\"ahler fields or represent the position of a brane 
in an internal space. In such models a non-trivial field-space metric $G_{AB}$,
with $A$ and $B$ being generic field-space indices, is naturally 
introduced (see for example \cite{Lyth:1998xn} for a review).
It is known \cite{Langlois:2008qf,Arroja:2008yy} that the non-Gaussianity of 
a multiple-field extension of DBI 
inflation where a probe D-brane moves in a multi-dimensional throat
can be studied using a metric $G_{AB}$ in combination with general kinetic 
terms of the form that we will define in the next section 
(for related works, see
\cite{Langlois:2008wt,
RenauxPetel:2008gi,Langlois:2009ej,
Gao:2009gd,Mizuno:2009cv,Gao:2009at,
Mizuno:2009mv,RenauxPetel:2009sj,Gong:2011uw,
RenauxPetel:2011uk,
McAllister:2012am,Kidani:2012jp,
Naruko:2012fe,
Fasiello:2013dla,Kobayashi:2013ina}).\footnote{It is worth noting that there 
is another type of multiple-field extension of DBI inflation where multiple 
branes move in different throats 
\cite{Cai:2008if,Cai:2009hw,Pi:2011tv,Emery:2012sm,Emery:2013yua}.
In this type of model, there is no natural field-space metric, and depending
on their specific Lagrangians and field trajectories they might not be of 
the type studied in this paper, as explained later on.}
These models are more general than the ones considered in for example
\cite{Langlois:2008mn,Gao:2008dt,Elliston:2012ab,Kaiser:2012ak,White:2013ufa} 
that depend on $X$ only (defined in the next section).
But again, since the analysis of \cite{Langlois:2008qf,Arroja:2008yy}
is based on the third-order action in terms of the fields like that of 
\cite{Seery:2005gb}, to compute the non-Gaussianity of $\zeta$ 
the $\delta N$ formalism is necessary, with its implied slow-roll approximation.
Here we go beyond that restriction.

In this paper we will derive the full exact
third-order action for what we call the diagonal class of models. 
This is a very general class of models with generalized kinetic terms
as mentioned above and defined in the next section, with the sole restriction 
being that the kinetic term is only coupled to tensors that are diagonal 
in the adiabatic-entropic field-space basis. 
This action is valid beyond any slow-roll or 
long-wavelength approximations. Our expressions are all written in a
manifestly covariant way. We also address the issue of the commutation
of covariant field derivatives and derivatives with respect to the 
kinetic term, showing that in a proper covariant framework they do actually
commute, which was not realized before in the literature. We define
a new hierarchy of slow-roll-like parameters, that describe all types
of derivatives of the Langrangian in a unified way. While most of these 
parameters will be small in the slow-roll approximation, no such 
approximation is made in this paper, all expressions are exact. 

The paper is organized as follows.   In Sec.~\ref{form} we introduce 
the basic elements of our formalism and discuss the issue of covariance. 
In Sec.~\ref{diag} we introduce the  diagonal class of models 
that describes the most general type of Lagrangians (within our starting 
assumptions) that allows for well-defined adiabatic and entropic sound speeds, 
and we introduce the new hierarchy of parameters.
For this diagonal class of models we provide the second-order and third-order
action for the adiabatic and isocurvature perturbations in 
Sec.~\ref{sec.second} and \ref{sec.third}, respectively. 
Next, in Sec.~\ref{sec.applications}, we apply our general results to the 
long-wavelength limit and to two-field DBI inflation. 
Finally, we conclude in Sec.~\ref{sec.conc}, followed by two  appendices
with additional details about some of the longer calculations in
the paper.

\section{Formalism}\label{form}

In this section we introduce the basic elements of our formalism,
which consist of the quantities we use to describe the generalized
inflationary background as well as the cosmological perturbations on
it. We will extend the results of \cite{Tzavara:2011hn} to accommodate
the new properties of generalized models of inflation.

\subsection{Background}\label{back}

We start here by defining our generalized background and exploring its
properties. In such a background one encounters two types of
derivatives: derivatives with respect to the kinetic terms and
covariant derivatives with respect to the fields. Hence it is crucial
to examine and understand their commutation. Furthermore, we study two
different field bases that are relevant to our models and are induced
by the dynamics of the system. Finally, we start building a hierarchy
of slow-roll parameters, a procedure that will be completed in 
section~\ref{diag}.

\subsubsection{Background dynamics}

We will consider an inflationary universe dominated by two scalar
fields, which in the background are homogeneous and will be denoted there by
$\phi^A$, with $A=1,2$. A generalization to any number of fields is
straightforward although calculations can be a lot more
cumbersome. The matter Lagrangian $P[X^{AB},G_{AB}(\phi),\phi^C]$ is a general 
function of the
kinetic term of the fields $X^{AB}$, the field metric $G_{AB}$, which
is taken to be an arbitrary function of the fields only, and the fields
themselves. 
The action that describes the physical system of
space-time and matter is
\be
S=\frac{1}{2}\int\d ^4x\sqrt{-g}\left[\kappa^{-2}R+2P
\right], 
\label{starting_action}
\ee
where $\kappa^2\equiv8\pi G$ and $X^{AB}$ in the background reduces to
\be
X^{AB}=\frac{1}{2}\frac{\dphi^A}{N}\frac{\dphi^B}{N}.\label{xabback} 
\ee
It should be noted that, while quite general, $P$ does not describe all possible
types of kinetic terms: we do not consider higher derivative terms  (e.g.\ terms of the type $\phi\Box\phi$ are not covered
by this definition).
For the canonical case $P=X-W$, where
\be
X= G_{AB}X^{AB}=\frac{1}{2}\frac{\dphi^2}{N^2}
\ee
and $W$ is the field potential. We denote the length of $\dphi^A$ by $\dphi$,
defining the length of a vector as its norm:
\be
A\equiv |A^B|\equiv \sqrt{A_BA^B}.
\ee
Field indices are lowered and raised using the field metric $G_{AB}$ in the
usual way:
\be
A_B=G_{BC}A^C .
\ee
Furthermore, in (\ref{starting_action}) $R$ is the Ricci scalar and $g$ is 
the determinant of the Friedmann-Robertson-Walker metric,
\be
g_{\mu\nu}=-N^2\d t^2+h_{ij}\d x^i\d x^j, 
\ee
where $h_{ij}$ is the metric of the spatial hypersurfaces that is assumed to
be $a^2$ times the identity matrix, i.e.\ a flat universe. 
The rate of change of the scale factor $a$ is the Hubble parameter,
\be
H=\frac{1}{N}\frac{\dot{a}}{a},
\ee
a measure of the expansion rate of the universe. Here we have not made a choice
of time coordinate. $N=1$ corresponds to the usual cosmic time,
$N=a$ to the conformal time, while $N=1/H$ corresponds to a time
variable that coincides with the number of e-foldings.

The canonical conjugate momentum $\Pi_A$ of the fields will turn out to be an 
important quantity for our formalism. It is at the basis of the definition 
of the slow-roll parameters and the basis vectors in field-space, as explained 
later on. It is given by
\be
\Pi_A\equiv N\frac{\partial P}{\partial\dphi^A}=P_{\langle AB\rangle}\frac{\dphi^B}{N}
,\label{momdef}
\ee
where we have introduced the short-hand notation
\be 
P_{\langle AB\rangle}\equiv\frac{\partial P}{\partial X^{AB}}.\label{pab} 
\ee
The angular brackets are used to distinguish $X^{AB}$ derivatives from field derivatives, as well as to denote the symmetric properties of the derivative, following the notation of \cite{Langlois:2008qf}. Notice that the field momentum is not necessarily parallel to the velocity of the fields, defined as
\be 
\tilde{\Pi}^A\equiv\frac{\dphi^A}{N}.
\ee
 
Within this framework, the background field equation is given by
\be
\cD_t\Pi^A+3NH\Pi^A-NP^{,A}=0. \label{fe}
\ee
Here we first encounter the covariant time derivative, which we define in (\ref{def_cov_muder}) in the next subsection. This derivative reduces simply to $\dot{\Pi}^A$ in the canonical case, and hence the field equation acquires its canonical form, found for example in \cite{Tzavara:2010ge}. 
Note that it is due to the use of the conjugate momentum and the covariant derivative that the form of this equation remains simple and reminiscent of the canonical case equation. We will further investigate the properties of covariant derivatives in the next subsection \ref{cov}, as well as their commutation with kinetic-term derivatives. Finally, the Einstein equations are
\be 
H^2
=\frac{\kappa^2}{3}\lh \Pi_A\tilde{\Pi}^A-P\rh\equiv\frac{\kappa^2}{3}\rho\qquad\mathrm{and}\qquad
\dot{H}=-\frac{\kappa^2}{2}N\Pi_A\tilde{\Pi}^A.\label{EE}
\ee

\subsubsection{Covariant derivatives}\label{cov}

As mentioned above, in this paper we treat the case of two-field inflation, 
where the two fields $\phi^A$ are considered the coordinates of a 
field manifold with a possibly non-trivial field metric $G_{AB}(\phi)$. 
We define the covariant field derivative (denoted by a semicolon, as opposed 
to the normal derivative, which is denoted by a comma) on this manifold in 
the usual way ($\Gamma^A_{BC}$ is the Christoffel symbol defined from the field
metric $G_{AB}$ in the usual way):
\be
A^{AB\ldots}_{CD\ldots ;E} = A^{AB\ldots}_{CD\ldots ,E}
+ \Gamma^A_{EF} A^{FB\ldots}_{CD\ldots}
+ \Gamma^B_{EF} A^{AF\ldots}_{CD\ldots} + \ldots
- \Gamma^F_{CE} A^{AB\ldots}_{FD\ldots}
- \Gamma^F_{DE} A^{AB\ldots}_{CF\ldots} - \ldots.
\ee
Its purpose is to ensure that, when computing the derivative, the quantity
$A$ is not just transported in field space by $\delta\varphi^E$, but
transported in a parallel way (otherwise there is no physical meaning
of the derivative). Hence the covariant derivative is the natural
derivative on curved field spaces.

We express everything in a fully covariant way with respect to the
field manifold.  This has several advantages. In the first place, the
explicitly covariant equations are manifestly invariant under changes
of field coordinates.  Secondly, the explicitly covariant equations
are in general shorter and easier to understand. Finally, the fact
that the covariant derivative of the field metric is zero by
construction means that the action of raising and lowering field
indices commutes with taking covariant derivatives. That simplifies
several manipulations. For example, the quantity $A_A(\phi) A^A(\phi)$
is a scalar and hence the normal derivative and covariant derivative
are equal: $(A_A A^A)_{;C} = (A_A A^A)_{,C}$.
However, only with the covariant derivative can this be simplified
to $2 A_A A^A_{;C}$, it is {\em not} equal to $2 A_A A^A_{,C}$.

Working within the Lagrangian formulation of the theory in curved field 
space, the covariant derivative of the generalised velocity of the fields 
is zero, $\tilde{\Pi}_{A;B}=0$, since the generalised coordinates and 
velocities are independent. This means that
\be
X_{AB;C} = 0.
\label{condXABCzero}
\ee
As we will show, this is an important relation for making sense of
covariant derivatives in the presence of derivatives with respect to
both $\phi^A$ and $X^{AB}$.   In the only previous work that used a covariant
framework to describe the third-order action in inflation with generalized
kinetic terms, \cite{Gong:2011uw}, the authors implicitly 
assumed the non-covariant condition $X_{AB,C}=0$ even in a covariant 
framework on a curved field manifold, but that led to several
issues with in particular mixed covariant derivatives. In a covariant 
framework the relation (\ref{condXABCzero}) is
the natural way to express the independence of the generalised coordinates 
and velocities that live in a curved field space.  

The Lagrangian $P$ we are considering is a function of the fields $\phi^A$
and the field velocities as encoded in $X^{AB}$. (The field metric $G_{AB}$
is a function of the fields and does not need to be considered
separately for this argument.) Hence in our calculations we will typically
encounter derivatives with respect to $\phi^A$, which should be covariant
derivatives in our covariant formalism, and derivatives with respect to
$X^{AB}$, which are ordinary derivatives since $X^{AB}$ is a tensor living
in the tangent space. The question now is whether these two types of
derivatives can be exchanged in a mixed derivative, i.e.\ whether they
commute. In \cite{Gong:2011uw}, assuming implicitly that $X_{AB,C}=0$, 
the authors came to the conclusion that they do not commute. Here, however, 
we show that with the relation (\ref{condXABCzero}) they do commute.

To show this, we rewrite the Lagrangian $P$ as a function of some number
of contractions $z_i$ of functions $f_i^\lambda$ of $\phi^A$ and functions 
$g_{i\,\lambda}$ of $X^{AB}$:
\be
P(X^{AB}, \phi^C) = P(z_i) \equiv P(f_i^\lambda(\phi) g_{i\,\lambda}(X)).
\ee
Here $\lambda$ stands for any number of upper and lower indices $A$.\footnote{
As a random example, one could have the Lagrangian
\be
P = A_{AB} X^{BC} X_{CD} X^{DA} \exp[G_{EF} X^{EF}/(A_{GH} A^{GH} + X_{GH} X^{GH})],
\ee
where $A_{AB}$ is some function of the $\phi^A$ only.
In that case $P$ would be the function 
$P(z_1,z_2,z_3,z_4) = z_1 \exp[z_2/(z_3+z_4)]$
of 4 contraction variables of the above type, with
$f_{1\,AB} = A_{AB}$, $g_1^{AB} = X^{BC} X_{CD} X^{DA}$,
$f_{2\,AB} = G_{AB}$, $g_2^{AB} = X^{AB}$, $f_{3} = A_{AB} A^{AB}$, $g_3 = 1$,
$f_4 = 1$, and $g_4 = X_{AB} X^{AB}$.} 
Since $P$ is a scalar, $P_{,C} = P_{;C}$, and with the relation
(\ref{condXABCzero}) we find 
$P_{;C} = \sum_i (\der P / \der z_i) f^\lambda_{i\, ;C} g_{i\,\lambda}$ and hence
\be
P_{;C\langle AB \rangle} = \sum_j \sum_i \frac{\der^2 P}{\der z_i \der z_j} 
f^{\tilde{\lambda}}_{j} g_{j\,\tilde{\lambda}\,\langle AB \rangle} f^\lambda_{i\, ;C} g_{i\,\lambda}
+ \sum_i \frac{\der P}{\der z_i} f^\lambda_{i\, ;C} g_{i\,\lambda\,\langle AB \rangle}.
\label{tempPCAB}
\ee
Similarly, $P_{\langle AB \rangle} = \sum_i (\der P / \der z_i) f^\lambda_{i} 
g_{i\,\lambda\,\langle AB \rangle}$,
so that $P_{\langle AB \rangle ;C}$ is also equal to the right-hand side of 
(\ref{tempPCAB}). 
In the end this means that the covariant derivative with
respect to $\phi^A$ and the derivative with respect to $X^{AB}$ commute:
\be
P_{;C\langle AB \rangle} = P_{\langle AB \rangle ;C}.
\ee
Note that the fact that $P$ is a scalar does not matter, one can easily
see that the argument can be generalized to non-scalar quantities,
e.g.\ $P_{\langle AB \rangle ;C \langle DE \rangle} = P_{\langle AB \rangle \langle DE \rangle ;C}$.
For completeness' sake we add here that derivatives with respect to $X^{AB}$
also commute with each other, since they are normal derivatives, 
\be
P_{\langle AB \rangle \langle CD \rangle} \equiv 
\frac{1}{2} 
\left( \frac{\partial P_{\langle AB \rangle}}{\partial{X^{CD}}} + 
\frac{\partial P_{\langle AB \rangle}}{\partial{X^{DC}}}\right)
= P_{\langle CD \rangle \langle AB \rangle}\,,
\label{p_secondderiv_xab}
\ee
and that as usual two covariant derivatives with respect to $\phi^A$ do not
commute, but give terms proportional to the Riemann tensor:
\be
A^{A\ldots}_{B\ldots ;CD} - A^{A\ldots}_{B\ldots ;DC}
= - R^A_{\;ECD} A^{E\ldots}_{B\ldots} - \ldots
+ R^E_{\;BCD} A^{A\ldots}_{E\ldots} + \ldots
\label{covf}
\ee

Finally, we define the covariant space-time derivative $\cD_\mu$:
\be 
\cD_\mu\cQ^{\ldots}_{\ldots}=\cQ^{\ldots}_{\ldots;C}
\partial_\mu \phi^C + \cQ^{\ldots}_{\ldots\langle CD\rangle}\cD_\mu X^{CD}.
\label{def_cov_muder}
\ee
This is a derivative
with respect to the space-time coordinates $x^\mu$, but it is covariant with
respect to the field manifold.
In fact, the quantity $\cQ^{\ldots}_{\ldots}$ is a scalar with respect to 
space-time but can carry any number of field indices and derivatives with
respect to $\phi^A$ and $X^{AB}$. Note that the relation $X_{AB;C} = 0$ is
used in this definition.

As an example we can consider the covariant time derivative of 
$P_{\langle AB \rangle}$, which according to the definition above is given by
\be
\cD_t P_{\langle AB \rangle} = P_{\langle AB \rangle ;C} \dot{\phi}^C 
+ P_{\langle AB \rangle \langle CD \rangle} \cD_t X^{CD}.
\ee
One might worry that this expression is not compatible with the standard
definition of the covariant time derivative,
\bea
\cD_t P_{\langle AB \rangle} & = & \dot{P}_{\langle AB \rangle} 
- \Gamma^D_{AC} P_{\langle DB \rangle} \dot{\phi}^C
- \Gamma^D_{BC} P_{\langle AD \rangle} \dot{\phi}^C \nn\\
& = & P_{\langle AB \rangle ,C} \dot{\phi}^C 
+ P_{\langle AB \rangle \langle CD \rangle} \dot{X}^{CD}
- \Gamma^C_{AD} P_{\langle CB \rangle} \dot{\phi}^D
- \Gamma^C_{BD} P_{\langle AC \rangle} \dot{\phi}^D,
\eea
since one does not seem to have enough $\Gamma$ terms to make both
$P_{\langle AB \rangle ,C}$ and $\dot{X}^{CD}$ covariant, and in
particular one appears to end up with a covariant $P_{\langle AB  \rangle ;C}$ 
but a non-covariant $\dot{X}^{CD}$. However, this is
just a consequence of not keeping proper track of the free indices in
$P_{\langle AB \rangle}$. Working with an explicit example,
e.g.\ $P_{\langle AB \rangle} = f(\phi) X_{AB}$, one sees that there
is no need to `use' the $\Gamma$ terms for $f(\phi)$, since it is a
scalar with all indices internally contracted, but that on the
other hand the $\Gamma$ terms are exactly what is needed to construct
the covariant $\cD_t X_{AB}$. Completely generally, $P_{\langle AB \rangle}$
has by construction only two free indices, and the two $\Gamma$ terms will
be used to make the corresponding quantity covariant (be it an $X_{AB}$ or
something depending on $\phi^A$), while all other 
quantities have their indices internally contracted and do not need
additional $\Gamma$ terms to be written in a covariant way. Of course
this argument holds for quantities with any number of free indices, since
there is always one $\Gamma$ term per free index.

In order to work out (\ref{def_cov_muder}) further, one needs to have
an expression for $\cD_\mu X^{AB}$. It turns out that this is not
trivial. Examining equation (\ref{fe}) one can see why it is not
straightforward to compute $\cD_tX^{AB}$: (\ref{fe}) is given in
terms of the momentum of the fields and not their velocity. These two
are related as
\be
\Pi_A=P_{\langle AB\rangle}\tilde{\Pi}^B. \label{momvec}
\ee
In general, this relation cannot be solved explicitly for $\tilde{\Pi}^A$.

\subsubsection{Orthonormal basis}\label{ob}

Since we are studying two-field inflation, the field space is two-dimensional
and we need to define a basis. It will turn out to be very useful to define
a basis induced by the momentum vector $\Pi^A$, with the unit vectors
\be
e_1^A\equiv\frac{\Pi^A}{\Pi},\qquad
e_2^A\equiv\frac{\cD_t\Pi^A-e_1^Ae_{1B}\cD_t\Pi^B}
{|\cD_t\Pi^A-e_1^Ae_{1B}\cD_t\Pi^B|},\label{ebasis}
\ee
the first one parallel to the field momentum and the second one parallel to the
component of the rate of change of the field momentum perpendicular to the 
momentum itself. 
As will be explained in more detail in section~\ref{gip}, this choice is
motivated by the two types of scalar perturbations present in the system:
the adiabatic perturbation, which is a linear combination of the curvature 
and the energy density perturbation and corresponds to the $e_1$ direction, 
and the isocurvature or entropic perturbation, which is defined through the 
non-adiabatic pressure perturbation and which on super-horizon scales becomes 
orthogonal to the adiabatic one. 

One could in principle just as well choose to work with a field basis defined through the velocity of the fields $\tilde{\Pi}^A$ instead of the momentum:
\be 
\tilde{e}_{1}^A=\frac{\tilde{\Pi}^A}{\tilde{\Pi}}. 
\label{velobasis}
\ee
In the case of canonical kinetic terms these two bases coincide, since then there is no distinction between the momentum and velocity vectors, but this is no longer the case in general with generalized kinetic terms. 
The energy density $\rho$, along the perturbation of which we define the adiabatic direction in the flat gauge, depends by definition both on the velocity of the field $\tilde{\Pi}^A$ and on its momentum $\Pi^A$ (see equation (\ref{EE})), so the choice seems to be ambiguous. However, the reason for and more discussion about our particular choice of basis and the implications thereof follow in subsections \ref{gip} and \ref{lder}.

The vectors of the orthonormal basis obey 
\be
G_{AB}e_m^Ae_n^B=\delta_{mn},
\ee
which implies
\be
G_{AB}=e_{1A}e_{1B}+e_{2A}e_{2B},\label{gdec}
\ee  
where the indices $m,n$ can take the values $1$ or $2$ and shall from now on denote adiabatic and  isocurvature component, respectively (sometimes also
denoted as $\parallel$ and $\perp$, respectively). 

\subsubsection{Slow-roll parameters}

We start constructing the background slow-roll parameters by considering 
the $\e$ parameter,
\be
\e=-\frac{\dot{H}}{NH^2}=
\frac{\kappa^2P_{\langle AB\rangle}X^{AB}}{H^2}=\frac{\kappa^2\Pi_A\tilde{\Pi}^A}{2H^2}
=\frac{\kappa^2\Pi^2}{2\Xi H^2}.\label{edef}
\ee
In the last equation we have defined the parameter $\Xi$ that is equal to the ratio of the momentum to the velocity of the fields:
\be
\Xi\equiv\frac{\Pi^2}{\Pi_A\tilde{\Pi}^A}=\frac{\Pi^2}{2P_{\langle AB\rangle} X^{AB}}
=\frac{G^{AC}P_{\langle AB\rangle}P_{\langle CD\rangle} X^{BD}}{P_{\langle AB\rangle} X^{AB}}
. \label{xidef}
\ee 
In addition to $\e$, we construct a hierarchy of $\eta$ slow-roll parameters:
\be
\eta^{(n)}_A\equiv\frac{1}{H^{n-1}\Pi}\lh\frac{1}{N}\cD_t\rh^{n-1}\Pi_A,\label{ordsr}
\ee
i.e.\ considering derivatives of the field equation (\ref{fe}). 
$\eta^{(1)}_A$ is just $e_{1A}$. For $\eta^{(2)}_A$ we will use the short-hand notation 
$\eta_A$. 
We find 
\bea
&&\eta^A=\frac{\cD_t\Pi^A}{NH\Pi}=-3e_{1}^A+\frac{1}{H\Pi}P^{,A},\nn\\
&&\hpa\equiv \eta^A e_{1A}=-3+\frac{1}{H\Pi}e_{1A}P^{,A},\qquad 
\hpe\equiv \eta^A e_{2A}=\frac{1}{H\Pi}e_{2A}P^{,A}.
\eea
These equations are identical to the definitions of $\eta^\parallel$ and $\eta^\perp$ for the canonical kinetic term case. This is due to expressing the field equation in terms of the conjugate momentum $\Pi^A$ as well as choosing to define the field basis through the same quantity.  
For the second order slow-roll parameters we find
\bea
&&\xi_A\equiv\eta^{(3)}_A=\frac{1}{NH^2\Pi}\cD_t\lh\frac{\cD_t\Pi_A}{N}\rh
=3\lh\e e_{1A}-\eta_A\rh+\frac{1}{NH^2\Pi}\cD_tP_{,A},
\nn\\
&&\xi^\parallel=3(\e-\hpa)+\frac{1}{NH^2\Pi}\cD_tP_{,A}e_1^A,\qquad \xi^\perp=-3\hpe+\frac{1}{NH^2\Pi}\cD_tP_{,A}e_2^A,
\eea
and so on.

For the time derivatives of the slow-roll parameters and basis vectors 
it is straightforward to find
\bea
&&\dot{\e}=2NH \e \lh \e+\hpa+\frac{\dot{\Xi}}{2NH \Xi}\rh\qquad\cD_te_1^A=NH\hpe e_2^A,\qquad\cD_te_2^A=-NH\hpe e_1^A\nn\\
&&\cD_t\eta^A=NH(\xi^A+\e\eta^A-\hpa\eta^A)\nn\\
&&\dot{\eta}^\parallel=NH\lh\xi^\parallel+\e\hpa+(\hpe)^2-(\hpa)^2\rh,\qquad
\dot{\eta}^\perp=NH\lh\xi^\perp+\e\hpe-2\hpa\hpe\rh
\label{derslow}\\
&&\cD_t\xi^A=NH\lh\eta^{(4)A}+(2\e-\hpa)\xi^A\rh\nn\\
&&\dot{\xi}^\parallel=NH\lh\eta^{(4)\parallel}+(2\e-\hpa)\xi^\parallel+\hpe
\xi^\perp\rh,\qquad
\dot{\xi}^\perp=NH\lh\eta^{(4)\perp}+(2\e-\hpa)\xi^\perp-\hpe
\xi^\parallel\rh.\nn
\eea
All these expressions, except for $\dot{\e}$, are identical to the ones with canonical kinetic terms. 

Although we defined the slow-roll parameters in terms of $\Pi^A$, one could in
principle also define slow-roll parameters through the velocity of the fields $\tilde{\Pi}^A$:
\be
\tilde{\eta}^{(n)}_A\equiv\frac{1}{H^{n-1}\tilde{\Pi}}\lh\frac{1}{N}\cD_t\rh^{n-1}\tilde{\Pi}_A.\label{tildeh}
\ee
For some expressions later on it will be useful to have also defined these
$\tilde{\eta}$ quantities. Note again that for canonical kinetic terms there
is no difference between $\eta$ and $\tilde{\eta}$.

Under the normal assumption that the derivatives of slow-roll parameters should
be one order higher in slow roll, we find from the equation for $\dot{\e}$
that in the non-canonical case there exists an additional slow-roll parameter,
namely $\dot{\Xi}/(NH\Xi)$. One can rewrite it as follows:
\be 
\frac{\dot{\Xi}}{\Xi NH}=2\hpa-\frac{1}{NHP_{\langle AB\rangle}X^{AB}}
\Bigg[
\lh P_{\langle AB\rangle\langle CD\rangle}
\mathcal{D}_tX^{CD}
+P_{\langle AB\rangle ;E}\dphi^E
\rh  X^{AB}
+
P_{\langle AB\rangle}\mathcal{D}_tX^{AB}
\Bigg].\label{eddef}
\ee
We cannot further simplify this expression for the time being, since we do not have an expression for $\mathcal{D}_tX^{AB}$. We shall do so in section \ref{diag}.

\subsection{Perturbations}

In this subsection we introduce the tools to study perturbations on
the generalized background. We start by reviewing the ADM
formalism. Next we define the adiabatic and isocurvature
perturbations using the field-space basis (\ref{ebasis}), generalizing the
work of \cite{Tzavara:2011hn}. We finally study the sound speeds
relevant in generalized models of inflation.

\subsubsection{ADM setup}

In order to study perturbations on the background described in subsection \ref{back}, we will work in the context of the ADM formalism and write the fully non-linear metric $\bg_{\mu\nu}$ as
\be
ds^2=-\bN^2dt^2+\bh_{ij}(dx^i+N^idt)(dx^j+N^jdt),\label{metricexact}
\ee
where $\bN$ is the lapse function and $N^i$ the shift. From now on
barred quantities will denote the fully non-linear quantity,
as opposed to their non-barred version that denotes the background part. 
For quantities that do not have a background part, as for example
$N^i$, there is no need to introduce the bar and in general we will not
write it to lighten the notation. Note that for the field the fully nonlinear 
version is distinguished by the use of the 
symbol $\varphi^A$, with the the background field being denoted $\phi^A$ as 
in the previous section.
The action takes the form
\be
\bS=\frac{1}{2}\int\d^4x\sqrt{\bh}\bN\lh\kappa^{-2}\bR^{(3)}+2\bar{P}(\bar{X}^{AB},\bG_{AB},\varphi^I)\rh 
+\frac{1}{2}\int\d^4x\sqrt{\bh}\bN^{-1}\kappa^{-2}\lh \bE_{ij}\bE^{ij}-\bE^2\rh,\label{actionexact}
\ee
where $\bh$ is the determinant of the space metric $\bh_{ij}$. For the purposes of this paper, we will consider a spatial metric of the form
\be
\bh_{ij}=a^2\mathrm{e}^{2\alpha}\delta_{ij}, 
\ee
where $\alpha$ is the curvature perturbation. 
$\bR^{(3)}$ is the intrinsic 
3-curvature, and the tensor $\bE_{ij}$ (proportional to the extrinsic curvature $\bK_{ij}=-\bN^{-1}\bE_{ij}$) is
\be
\bE_{ij}=\frac{1}{2}\left(\dot{\bh}_{ij}-D_i N_j-D_j N_i\right).
\ee
$D_i$ is the covariant space-time derivative on the space hypersurface, 
being just $\partial_i$ when acting on space scalars. Indices $i,j,k$ will 
denote space coordinates.

The fully non-linear form of the kinetic term reads
\be
\bar{X}^{AB}=-\frac{1}{2}\bg^{\mu\nu}\partial_{\mu}\varphi^A
\partial_{\nu}\varphi^B
=\frac{1}{2}\Big[\frac{1}{\bN^2}\lh\dot{\varphi}^A-N^j\partial_j\varphi^A\rh\lh
\dot{\varphi}^B-N^j\partial_j\varphi^B\rh
-\bh^{ij}\partial_i\varphi^A
\partial_j\varphi^B\Big],
\ee 
while the canonical momentum of the fields is generalised to 
\be
\bar{\Pi}_A\equiv \bN\frac{\partial \bP}{\partial\dot{\varphi}^A}
=\bP_{\langle AB\rangle}\frac{\dot{\varphi}^B-N^j\partial_j\varphi^B}{\bN}.
\ee
Minimising the action with respect to the metric elements gives two constraint equations for the system: the $N_i$ momentum constraint, 
\be
D_j\Big[\frac{1}{\bar{N}}(\bE_i^j-\bE\delta_i^j)\Big]=\kappa^2\bar{\Pi}_A
\partial_i\varphi^A,\label{mom}
\ee
and the $\bN$ energy constraint,
\be
\kappa^{-2}\bR^{(3)}+2\bP-2\bP_{\langle AB\rangle}\lh2\bX^{AB}+\bh^{ij}\partial_i
\varphi^A\partial_j\varphi^B\rh
-\bN^{-2}\kappa^{-2}\lh \bE_{ij}\bE^{ij}-\bE^2\rh=0.\label{en} 
\ee

\subsubsection{Gauge-invariant perturbations}\label{gip}

Any fully non-linear quantity can be expanded as an infinite series of perturbations as
\be
\bar{\mathcal{Q}}=\mathcal{Q}+\mathcal{Q}_{(1)}+\frac{1}{2}\mathcal{Q}_{(2)}+\cdots 
\ee
We choose the adiabatic perturbation at first order to be in the direction of the energy density perturbation in the flat gauge (see e.g.~\cite{Tzavara:2011hn}). 
Its general form is 
\be
\zeta_{1}\equiv\alpha_{(1)}-\frac{NH}{\dot{\rho}}\rho_{(1)}.
\ee
From (\ref{EE}) we find that for the background
\be
\dot{\rho}=-3H\Pi_A\dot{\phi}^A=-6NHP_{\langle AB\rangle}X^{AB}.
\ee
By expanding to first order the momentum constraint (\ref{mom}), 
we find that
\be
H_{(1)}\equiv\frac{\kappa^2}{6H}\rho_{(1)}=-\frac{\kappa^2}{2}\Pi_A\varphi^A_{(1)}.
\ee
Hence we finally get 
\be
\zeta_{1}=\alpha_{(1)}-H\frac{N\Pi_A}{\Pi_B\dot{\phi}^B}\varphi^A_{(1)}=\alpha_{(1)}-\frac{H\Xi}{\Pi}\frac{\Pi_A}{\Pi}\varphi^A_{(1)}=
\alpha_{(1)}-\kappa\sqrt{\frac{\Xi}{2\e}}e_{1A}
\varphi^A_{(1)}.\label{adia}
\ee
The vector coupled to the perturbation of the field is the momentum $\Pi^A$ and hence it is indeed the momentum that we have to use in order to define the proper field basis to discriminate between adiabatic and isocurvature effects.

The orthogonal combination, which is proportional to the isocurvature 
perturbation outside the horizon,\footnote{Even though strictly speaking inside
the horizon the isocurvature perturbation as defined through the pressure and
energy density perturbations, $p_{(1)}-(\dot{p}/\dot{\rho})\rho_{(1)}$, see
\cite{Wands:2000dp}, is not proportional to just $\zeta_2$ but also has a term proportional
to $\dot{\zeta}_1$ (the latter being proportional to $\zeta_2$ on super-horizon
scales), we will still refer to this orthogonal combination as the isocurvature
perturbation on all scales. For a related study of these gauge-invariant quantities in terms of the energy density and pressure perturbations within generalized inflationary models, see \cite{karimgen}.} 
takes the form
\be
\zeta_2 =-\frac{H\Xi}{\Pi}e_{2A}\varphi^A_{(1)}=-\kappa\sqrt{\frac{\Xi}{2\e}}e_{2A}
\varphi^A_{(1)}.\label{iso}
\ee
Hence, we have expressed the adiabatic and isocurvature perturbations in terms of the field perturbations, in agreement with our previous work \cite{Tzavara:2011hn}.

\subsubsection{Defining sound speeds}

We shall finish this section by considering the implications of a non-trivial kinetic term on the kinematics of the perturbations. This results in a non-trivial sound speed for the perturbations, manifesting itself as a ratio of the gradient to the time derivative of the perturbations different from unity.  We will define the adiabatic and entropic sound speed in terms of the kinetic term and gradient term of the second order action of the field perturbation in the flat gauge. It can be found by perturbing the action (\ref{actionexact}) to second order,
keeping here only the time and space derivative terms (indicated by the
label $(\textrm{k-g})$): 
\begin{eqnarray}
S_{(2)} ^{(\textrm{k-g})} = \frac{1}{2} \int \d t \d^3 x \, a^3 
\left[(P_{\langle AB \rangle}  + 2P_{\langle AC \rangle\langle BD 
\rangle } X^{CD}) \mathcal{D}_t\varphi^A \mathcal{D}_t\varphi^B - P_{\langle AB \rangle} 
h^{ij} \mathcal{D}_i\varphi^A \mathcal{D}_j \varphi^B\right]\,.
\label{second_order_action_main}
\end{eqnarray} 
Here $\varphi^A$ denotes the first order perturbation of the fields, we have suppressed the subscript $(1)$ in order to keep notation light.

In order to study the properties of these perturbations, we decompose them into adiabatic and entropy components with the help of the basis defined in (\ref{ebasis}),
\begin{eqnarray}
\varphi^A = \varphi_m e^A _{\;m}\,.
\end{eqnarray}
In terms of the decomposed fields, 
(\ref{second_order_action_main}) can be rewritten as
\begin{eqnarray}
S_{(2)} ^{(\textrm{k-g})} = \frac{1}{2} \int \d t \d^3 x \, a^3 
\left[\mathcal{K}_{mn}\dot{\varphi}_m\dot{\varphi}_n 
-\mathcal{G}_{mn} h^{ij} \partial_i\varphi_m \partial_j\varphi_n\right]\,,
\label{second_order_action_main_aden}
\end{eqnarray} 
where
\begin{eqnarray}
\mathcal{K}_{mn}  &\equiv& 
(P_{\langle AB \rangle}  + 2P_{\langle AC \rangle\langle BD 
\rangle } X^{CD})) e^A _{\;m} e^B _{\;n}\,,\label{kmn}\\
\mathcal{G}_{mn}  &\equiv& P_{\langle AB \rangle} 
e^A _{\;m} e^B _{\;n}\,.\label{gmn}
\end{eqnarray} 
Note that the covariant derivatives have become ordinary ones, since $\varphi_m$ is a scalar in field space. 

In order to define an adiabatic and entropic sound speed, the matrices $\mathcal{K}_{mn}$ and $\mathcal{G}_{mn}$ should be diagonal. Then one takes the ratio of the coefficients of the time derivative and the gradient terms:
\be
\frac{1}{c_{ad}^2}\equiv\frac{\mathcal{K}_{11}}{\cG_{11}}\qquad\mathrm{and}\qquad\frac{1}{c_{en}^2}\equiv\frac{\cK_{22}}{\cG_{22}}.\label{sound} 
\ee
While one could in principle define sound speeds even if the matrices
are not diagonal in this basis (for example in a similar way as one
determines the eigenfrequencies of a system of coupled harmonic
oscillators), these would then not be the adiabatic and entropic sound
speeds. We restrict ourselves in the rest of the paper to those
models where $\mathcal{K}$ and $\mathcal{G}$ are diagonal in the
adiabatic-entropic basis, so that the sound speeds are the adiabatic and 
entropic ones. 
In the next section we will investigate the conditions
for this to be the case, and find that the resulting class of models allows
for important simplifications in the calculations.

\section{The diagonal class of models}\label{diag}

In this section we introduce the most general class of models that
allows for well-defined adiabatic and entropic sound speeds. After
demonstrating the necessary conditions to achieve that, we reexamine
the sound speeds as well as the slow-roll parameters. We complete our
study by defining generalized slow-roll parameters with mixed
kinetic-term and covariant-field derivatives.

\subsection{Kinetic-term derivatives}{\label{lder}}

In order to define the adiabatic and entropic sound speeds and hence
in order to study when the matrices $\mathcal{K}_{mn}$ and
$\mathcal{G}_{mn}$ are diagonal, we need to examine the properties of
the derivatives of the Lagrangian with respect to the
kinetic term.  We start with a general Lagrangian $\bP$ of the form
$\bP=\bP(\bX^{AB},\bG^{AB},\varphi^A)$, where we remind the reader that
the bar is used to indicate fully non-linear quantities, while $\varphi^A$
is also the fully non-linear field, its background version being denoted 
$\phi^A$. The derivative $\bP_{\langle AB\rangle}$ can be decomposed in the 
directions of an arbitrary orthonormal basis of unit vectors in the two-field 
space $\{\bar{\tilde{e}}_1,\bar{\tilde{e}}_2\}$ as
\be
\bP_{\langle AB\rangle}=\bP_{11}\bar{\tilde{e}}_{1A}
\bar{\tilde{e}}_{1B}+\bP_{12}
\lh\bar{\tilde{e}}_{1A}\bar{\tilde{e}}_{2B}
+\bar{\tilde{e}}_{2A}\bar{\tilde{e}}_{1B}\rh+
\bP_{22}\bar{\tilde{e}}_{2A}\bar{\tilde{e}}_{2B},
\ee
where we keep in mind that $\bP_{\langle AB\rangle}$ is by construction symmetric. 
Let us now set the tilded orthonormal basis 
$\{\bar{\tilde{e}}_1,\bar{\tilde{e}}_2\}$ through the relation 
\be
\bX^{AB}= \bX\bar{\tilde{e}}_1^A
\bar{\tilde{e}}_1^B+\bX'\bar{\tilde{e}}_2^A
\bar{\tilde{e}}_2^B,
\ee
i.e.\ the basis that diagonalizes the non-linear kinetic term (which can always
be done since $\bX^{AB}$ is symmetric), with $\bX$ and $\bX'$ its eigenvalues. 
Now one can rewrite the derivative of the Lagrangian with respect to the 
kinetic term as
\be
\bP_{\langle AB\rangle}=\bP_\bX \bX_{AB}+ \bP_{12}\lh\bar{\tilde{e}}_{1A}\bar{\tilde{e}}_{2B}
+\bar{\tilde{e}}_{2A}\bar{\tilde{e}}_{1B}\rh
+\bP_{\bG}\bG_{AB},
\label{PAB_PX_PG}
\ee
with $\bP_\bX$ and $\bP_\bG$ defined by the relations 
$\bP_{11}=\bP_\bX\bX+\bP_\bG$ and $\bP_{22}=\bP_\bX\bX'+\bP_\bG$. 
The determinant of $\bX^{AB}$ is in general non-zero. However, the determinant
of the background $X^{AB}$ is zero, meaning that one of the eigenvalues goes
to zero in the background limit. From the definition (\ref{xabback}) of the
background $X^{AB}$,
\be
X^{AB}\equiv\frac{1}{2}\frac{\dphi^A}{N}\frac{\dphi^B}{N}=X\tilde{e}_1^A\tilde{e}_1^B ,
\label{xdec}
\ee
we see that in the background the basis 
$\{\bar{\tilde{e}}_1,\bar{\tilde{e}}_2\}$
reduces to the velocity basis (\ref{velobasis}) so that our notation using
the tilde is consistent. Moreover, $\bX$ becomes $X$, while $\bX'$ goes to 
zero in the background.

In the background the momentum vector (\ref{momvec}) can be decomposed in the 
tilded basis as
\be
\Pi_A=\frac{\dphi}{N}\lh P_{11}\tilde{e}_{1A}+P_{12}\tilde{e}_{2A}
\rh.\label{vectors} 
\ee
It is clear that in the general case the momentum and the velocity vectors are 
not parallel and hence the definition of the most convenient and physical 
field-space basis is ambiguous. This can be resolved by setting $P_{12}=0$. 
In that case the momentum and velocity vectors are parallel and the background
momentum basis $\{e_1,e_2\}$ and velocity basis $\{\tilde{e}_1,\tilde{e}_2\}$
are identical and can be used interchangeably. 
We will in fact impose the same condition on the fully non-linear Lagrangian:
$\bP_{12}=0$. Then
\be 
\bP_{\langle AB\rangle}=\bP_{11}\bar{\tilde{e}}_{1A}
\bar{\tilde{e}}_{1B}+
\bP_{22}\bar{\tilde{e}}_{2A}\bar{\tilde{e}}_{2B}
= \bP_\bX \bX_{AB}+\bP_{\bG}\bG_{AB},
\label{new}
\ee 
where $\bP_{11},\ \bP_{22},\ \bP_\bX,$ and $\bP_\bG$ are functions of the fields 
$\varphi^A$ and their derivatives $\bX^{AB}$.
Demanding $\bP_{12}=0$ is equivalent to studying models where $\bX^{AB}$ is 
only contracted with tensors of the form
\be
\bA_{AB}=\bA_{11}\bar{\tilde{e}}_{1A}
\bar{\tilde{e}}_{1B}
+\bA_{22}\bar{\tilde{e}}_{2A}\bar{\tilde{e}}_{2B}, 
\ee
i.e.\ tensors $\bA_{AB}$ that are diagonal in the 
$\{\bar{\tilde{e}}_1,\bar{\tilde{e}}_2\}$ basis. 
Although restricted, this class of models encompasses a very wide range of 
Lagrangians. On the other hand, terms of the form $\varphi^A\varphi^B$ 
coupling to the kinetic term are excluded, since we would end up with 
additional mixed terms $\bar{\tilde{e}}_{1A}\bar{\tilde{e}}_{2B}$ as soon as 
the field trajectory makes a turn in field space. The above arguments  
remain true for higher-order kinetic-term derivatives: as long as $\bX_{AB}$ 
is only contracted with diagonal tensors, no mixed terms can appear. 
We show in the next section that the condition $\bP_{12}=0$ is sufficient
to have $\mathcal{K}_{12}=\mathcal{G}_{12}=0$. It also significantly simplifies
calculations.

Finally, since the two bases $e_{mA}$ and $\tilde{e}_{mA}$ coincide in the 
background, we find that for these models
\be
\Xi=P_{\bar{0}}=P_XX+P_G.\label{xp}
\ee
In order to lighten the notation, we introduced here barred indices, which 
from now on shall denote kinetic-term derivatives contracted with basis vectors,
in the following way:
\bea 
&&P_{\langle AB\rangle}e_1^Ae_1^B=P_{\bar{0}}\nonumber\\
&&P_{\langle AB\rangle}e_1^Ae_2^B=P_{\bar{1}}\nonumber\\
&&P_{\langle AB\rangle}e_2^Ae_2^B=P_{\bar{2}}.
\eea 
Note that although $P_{\bar{1}}=0$ for the diagonal models, the $\bar{1}$ 
notation is still required since second-order mixed derivatives, like 
$P_{\bar{1}\bar{1}}$, are not necessarily zero (remember that all derivatives 
with respect to $X^{AB}$ are taken first, before the contraction with the 
basis vectors, and that those two operations do not commute). 

\subsection{Sound speed revisited}

For the class of models described in the previous section the matrices 
$\mathcal{K}_{mn}$ and $\mathcal{G}_{mn}$ are diagonal. In order to see that, 
we compute the second derivative (\ref{p_secondderiv_xab})
\bea
\bP_{{\langle AB\rangle}{\langle CD\rangle}}
&=&\frac{\bP_\bX}{2}\lh \bG_{AC}\bG_{BD}+\bG_{AD}\bG_{BC}\rh
+\bP_{\bX\bX}\bX_{AB}\bX_{CD}+\bP_{\bG\bG}\bG_{AB}
\bG_{CD}\nn\\
&&+\frac{\bP_{\bX\bG}}{2}\lh \bG_{AB}\bX_{CD}+\bX_{AB}\bG_{CD}\rh 
,\label{sder}
\eea 
by differentiating (\ref{new}) with respect to $\bX^{CD}$. It now becomes 
clear why we chose to express $\bP_{\langle AB\rangle}$ in terms of $\bX_{AB}$ 
and $\bG_{AB}$ and not $\bar{\tilde{e}}_{mA}\bar{\tilde{e}}_{mB}$: the form 
of equation (\ref{new}) allows us to easily compute the second-order 
derivative with respect to $\bX^{CD}$. The first term in (\ref{sder}) comes 
from the derivative of $\bX^{AB}$, while the rest of the terms are related 
to the derivatives of $\bP_\bX$ and $\bP_\bG$:
\bea
\bP_{\bX {\langle CD\rangle}}
&=& \bP_{\bX\bX} \bX_{CD} + \frac12 \bP_{\bX\bG} \bG_{CD}
\nonumber\\
\bP_{\bG {\langle CD\rangle}}
&=& \frac12 \bP_{\bX\bG} \bX_{CD} + \bP_{\bG\bG} \bG_{CD}.
\label{PX_PG_ders}
\eea
The equations (\ref{PX_PG_ders}) just express the general fact that any 
quantity with indices $_{CD}$ can be expressed in the basis 
$\bar{\tilde{e}}_{1C}\bar{\tilde{e}}_{1D}$, 
$\bar{\tilde{e}}_{1C}\bar{\tilde{e}}_{2D}$,
$\bar{\tilde{e}}_{2C}\bar{\tilde{e}}_{1D}$, 
$\bar{\tilde{e}}_{2C}\bar{\tilde{e}}_{2D}$
and that the terms proportional to $\bar{\tilde{e}}_{1C}\bar{\tilde{e}}_{1D}$
and $\bar{\tilde{e}}_{2C}\bar{\tilde{e}}_{2D}$ can be replaced by terms
proportional to $\bX_{CD}$ and $\bG_{CD}$ like in (\ref{PAB_PX_PG}).
The quantities $\bP_{\bX\bX}$, $\bP_{\bX\bG}$, $\bP_{\bG\bG}$ are defined by 
these equations as the coefficients of the $\bX_{CD}$ and $\bG_{CD}$ components.
The absence of the cross terms  $\bar{\tilde{e}}_{1C}\bar{\tilde{e}}_{2D}$ and 
$\bar{\tilde{e}}_{2C}\bar{\tilde{e}}_{1D}$ in (\ref{PX_PG_ders}) is due to the 
symmetry $\bP_{{\langle AB\rangle}{\langle CD\rangle}} = 
\bP_{{\langle CD\rangle}{\langle AB\rangle}}$, and the same symmetry also imposes 
that $\bP_{\bG\bX}=\bP_{\bX\bG}$.

From the background version of (\ref{sder}) we now find that 
$P_{{\langle AC\rangle}{\langle BD\rangle}} X^{CD} e_1^A e_2^B = 0$ using the 
orthogonality of the basis as well as (\ref{gdec})
and (\ref{xdec}), which now reads $X^{AB}=Xe_1^Ae_1^B$. Since $P_{\bar{1}}=0$ we
conclude from (\ref{kmn}) and (\ref{gmn})
that $\mathcal{K}_{12}=\mathcal{G}_{12}=0$, so that these matrices are
indeed diagonal as claimed.
The background values of $\mathcal{K}_{mn}$ and $\mathcal{G}_{mn}$ take the form 
\bea
\mathcal{K}_{mn}=\mathrm{diag}\Big[P_{\bar{0}}+2XP_{\bar{0}\bar{0}},
P_{\bar{2}}+2XP_{\bar{1}\bar{1}}\Big]=
\mathrm{diag}\Big[\Xi+2XP_{\bar{0}\bar{0}},\Xi\Big]
\label{kmn2}
\eea
and
\be
\mathcal{G}_{mn}=\mathrm{diag}
\Big[P_{\bar{0}},P_{\bar{2}}\Big]
= \mathrm{diag}
\Big[\Xi,P_{\bar{2}}\Big].
\ee
To find the second equality in (\ref{kmn2}) we used (\ref{sder}) to get 
$P_{\bar{1}\bar{1}}=P_X/2$, as well as (\ref{xp}).
Using (\ref{sound}), this translates into an adiabatic and entropic sound speed
\bea
\frac{1}{c_{ad}^2}=1+\frac{2XP_{\bar{0}\bar{0}}}{\Xi}
\quad\mathrm{and}\quad
\frac{1}{c_{en}^2}=\frac{\Xi}{P_{\bar{2}}}\,.
\label{sound_speed_diagonal}
\eea 

\subsection{Slow-roll parameters revisited}

The fact that the background momentum and field-velocity field-space bases 
coincide in the
diagonal class of models allows for major computational simplifications.
Within these models we can compute $\cD_tX^{AB}$ (and hence further simplify expressions like (\ref{eddef})) as
\be 
\cD_tX^{AB}=\dot{X}e_1^Ae_1^B+2NHX\hpe(e_2^Ae_1^B+e_1^Ae_2^B),
\ee 
where we used (\ref{derslow}) for the derivatives of the unit vectors. Furthermore, the complicated relation between $\Xi$ and $X^{AB}$ (\ref{eddef}) is now replaced by the far more simple (\ref{xp}). The time derivative of $X$ itself is calculated a little further down in this section, in (\ref{tildedef}).  The rest of this section is devoted to exploring the consequences of this simplification for the background quantities.  

Now that we have set $e_{mA}=\tilde{e}_{mA}$,   
we can also introduce a hierarchy of new parameters that incorporates all types
of derivatives on $P$:
\be
\Lambda^{A_1\cdots A_p}
_{\bar{m}_1\cdots \bar{m}_q}
=
\frac{3^{\mathrm{sgn}(p)}}{\Xi}\frac{\lh 2X\rh^{\frac{p}{2}+q-1}}{\lh 3H\rh^p}
P^{;A_1;\cdots ;A_p}_{\langle B_1\tilde{B}_1\rangle\cdots \langle B_q\tilde{B}_q\rangle}
e_{m_1}^{B_1}e_{\tilde{m}_1}^{\tilde{B}_1}
\cdots 
e_{m_q}^{B_q}e_{\tilde{m}_q}^{\tilde{B}_q},
\label{newl}
\ee
where $\mathrm{sgn}(p)$ is the sign function, i.e.\ $\mathrm{sgn}(0)=0$ and 
$\mathrm{sgn}(p)=1$ for $p>0$. 
While definition (\ref{newl}) might appear complicated, it is actually a very convenient unified description of the possible derivatives on $P$. Whenever a quantity like $\Lambda^{AB}_{\bar{0}\bar{1}\bar{2}}$ appears in an expression, the reader will know that it means $2$ covariant field derivatives on $P$ (upper indices) and $3$ kinetic derivatives with respect to $X^{AB}$ (lower indices), the latter being respectively of pure adiabatic ($\bar{0}=11$), mixed ($\bar{1}=12$), and pure isocurvature ($\bar{2}=22$) type. Of course, in practice the upper
indices will also often be contracted with basis vectors.  
We remind the reader that first-order field derivatives ($p=1$) contracted with $e_{1A}$ are denoted by $\parallel$ and contracted with $e_{2A}$ by $\perp$.  
The prefactor of (\ref{newl}) is there just for dimensional reasons and because the derivatives in the equations always occur with these factors. Note that the $\Lambda^{A_1\cdots A_p}
_{\bar{m}_1\cdots \bar{m}_q}$ parameters are invariant under permutations of the lower indices.  On the other hand, they are not invariant under permutations of the upper indices, since covariant derivatives do not commute (see (\ref{covf})). This is strictly true for three or more upper indices; in the case of two field derivatives, the upper indices do commute, since one of them is a normal derivative. 

The ordinary slow-roll parameters (\ref{ordsr}) are  related to the new generalized parameters (\ref{newl}).
For example, for  $p=1$ and $p=2$ one can write
\bea 
&&\eta^A=-3e_{1}^A+\Lambda^{A},\nn\\
&&
\xi^A=3\lh\e e_1^A-\eta^A+\Lambda^{1A}\rh
+\tilde{\eta}^\parallel\Lambda^A_{\bar{0}},
\eea
and so on. However, for these previously defined normal slow-roll parameters we shall keep  the standard notation $\eta^{(n)A}$. In addition, the $\Lambda^{22}$ component is related to the first-order $\chi$ parameter, already known from the canonical case, 
\be
\chi \equiv -\Lambda^{22}+\frac{\dot{\e}}{2NH\e}.
\ee 
Notice that $\chi$ is not necessarily small during a slow-roll period, so in that sense it is not really a slow-roll parameter.
It is worth mentioning that the sound speeds are also related to the parameters
 (\ref{newl}), the ones with $p=0$ (pure kinetic derivatives):
 \bea
\frac{1}{c_{ad}^2}=1+\Lambda_{\bar{0}\bar{0}}
\quad\mathrm{and}\quad
\frac{1}{c_{en}^2}=\frac{1}{\Lambda_{\bar{2}}}\,.
\label{sound_speed_diagonal_ito_Lambda}
\eea

Differentiating (\ref{xp}) we get 
\be
\frac{\dot{\Xi}}{NH\Xi}=\hpa(1-c_{ad}^2)+\Lambda_{\bar{0}}^\parallel c_{ad}^2
\ee
and hence\footnote{
It is important to note that even though we now have $e_m=\tilde{e}_m$, this
is caused by the fact that $\Pi^A$ and $\tilde{\Pi}^A$ have the same direction,
not the same magnitude. Hence there is no reason why $\thpa$, defined from
$\tilde{\Pi}^A$, and $\hpa$, defined from $\Pi^A$ should be the same, and indeed
they are not in general as this equation shows.}
\be 
\thpa=\frac{\dot{X}}{2NHX}=c_{ad}^2(\hpa-\Lambda_{\bar{0}}^\parallel)
\qquad\mathrm{and}\qquad\thpe=\hpe,
\label{tildedef}
\ee
indicating that the non-canonical character of the kinetic terms affects only 
the adiabatic direction in the background.
Although we prefer showing our expressions in terms of the new generalized $\Lambda$ parameters, the combination  $\tilde{\eta}^\parallel$ seems to be a characteristic quantity for general inflation models and hence from now on we will use this short-hand notation when this combination appears. 
For the derivative of $\e$ (\ref{eddef}) one finds
\be 
\dot{\e}=2NH\e \lh\e+\frac{\hpa+\tilde{\eta}^\parallel}{2}\rh.  
\ee
Notice that since for the diagonal class of models $\Pi^A$ and $\tilde{\Pi}^A$ are parallel vectors, we can rewrite $\e$ in terms of the magnitude of these vectors as 
\be
\e=\frac{\kappa^2\Pi\tilde{\Pi}}{2H^2}.
\ee
Finally, we also give here the derivatives of the sound speeds:
\bea
&&\frac{\dot{c}_{ad}}{NHc_{ad}}
=-\frac{c_{ad}^2}{2}
\Bigg\{
\Lambda_{\bar{0}}^\parallel
+\Lambda_{\bar{0}\bar{0}}^\parallel
+\tilde{\eta}^\parallel
\lh \frac{4}{c_{ad}^2}-3+\Lambda_{\bar{0}\bar{0}\bar{0}}\rh 
 - \frac{\hpa}{c_{ad}^2} 
 \Bigg\}\,,
\nn\\ 
&&
\frac{\dot{c}_{en}}{NHc_{en}}=\frac{1}{2c_{en}^2}
\Bigg\{
\Lambda_{\bar{2}}^\parallel
+\tilde{\eta}^\parallel
\lh c_{en}^2+\Lambda_{\bar{2}\bar{0}}\rh 
 -\hpa c_{en}^2 
 \Bigg\}. \label{cdef}
\eea

The time derivatives of the sound speeds are considered as slow-roll parameters in the literature, in the sense that one can set 
\be 
\frac{\dot{c}_{ad}}{NHc_{ad}}\ll 1\quad\mathrm{and}\quad\frac{\dot{c}_{en}}{NHc_{en}}\ll 1,\label{csr}
\ee 
assuming slow-roll. 
Furthermore, within the slow-roll approximation all time derivatives of slow-roll parameters are one order smaller as compared to the slow-roll parameters themselves. Keeping these statements in mind, one should assume that the 
$\Lambda^{A_1\cdots A_p}_{\bar{m}_1\cdots\bar{m}_q}$ with $p\neq0$ behave as ordinary slow-roll parameters, in the sense that the higher is the number of field derivatives (i.e.\ $p$), the higher is the order of the slow-roll parameter. This is not strictly true for the pure isocurvature field-derivative components, as is already known for the case of $\chi$ (which is not necessarily small during slow-roll evolution). On the other hand, the $p=0$ parameters (the pure kinetic derivatives) are not necessarily small assuming slow roll. 
For example, no condition on the order of the $p=0$ parameters occurs from demanding 
the left-hand side of (\ref{cdef}) to be small, since the 
$\Lambda_{\bar{0}\bar{0}\bar{0}}$ and $\Lambda_{\bar{2}\bar{0}}$
terms within the parentheses are already multiplied by the slow-roll parameter
$\tilde{\eta}^\parallel$. On the other hand, the same relation shows that they
should be at most of order $\cO(1)$, in order not to make (\ref{cdef}) large
after all.

After this discussion of the behaviour of these parameters in the slow-roll 
approximation, it should be emphasized that no such approximation is made
in this paper. All expressions here are exact and the slow-roll parameters
can be large. Finally, we would like to remind the reader that no explicit 
expressions for the slow-roll parameters (both the normal and the new ones) would be possible (i.e.\ being able to replace time derivatives $\mathcal{D}_t$ by
derivatives with respect to the fields and kinetic term) unless the momentum and velocity bases were coinciding. Hence, we can now continue studying perturbations on this background, restricting ourselves to the diagonal class of models.

\section{Second-order action}\label{sec.second}

Having set the framework of the diagonal class of models, we are now ready to study perturbations in these models. We shall work in the flat gauge $\alpha_{(r)}=0$, where $r$ is the perturbation order, and hence the adiabatic and isocurvature perturbations defined respectively in (\ref{adia}) and (\ref{iso}) take the form (see also \cite{Tzavara:2011hn})
\be
\zeta_m=-\frac{H}{\tilde{\Pi}}e_{mA}\varphi^A=
-\frac{H\Xi}{\Pi}e_{mA}\varphi^A=
-\kappa\sqrt{\frac{\Xi}{2\e}}e_{mA}\varphi^A.\label{gaugezeta}
\ee
In order to study first-order perturbations one needs to calculate the second-order action. To that end, one needs to perturb the ADM action (\ref{actionexact}) to second order and rewrite the field contributions in terms of the gauge-invariant combinations (\ref{gaugezeta}). Afterwards one can replace the $\cD_t\varphi^A$ terms occurring in the action by the expression
\be
\dot{Z}_m\equiv-\kappa\sqrt{\frac{\Xi}{2\e}}e_{mA}\mathcal{D}_t\varphi^A,\label{zcap}
\ee
where by differentiating (\ref{gaugezeta}), one can find that
\bea
&&\dot{Z}_1=\dot{\zeta_1}
+\frac{\Xi}{2\e}\frac{\d }{\d t}\lh\frac{\e}{\Xi}\rh\zeta_1
-\hpe NH\zeta_2
=\dot{\zeta_1}
+\lh\e+\tilde{\eta}^\parallel\rh 
NH\zeta_1
-\hpe NH\zeta_2
\nn\\
&&\dot{Z}_2=\dot{\zeta_2}
+\frac{\Xi}{2\e}\frac{\d }{\d t}\lh\frac{\e}{\Xi}\rh\zeta_2
+\hpe NH\zeta_1
=\dot{\zeta_2}
+\lh\e+\tilde{\eta}^\parallel \rh NH\zeta_2
+\hpe NH\zeta_1.\nn\label{derzeta}
\eea
Finally, one needs to use the first-order momentum  constraint (\ref{mom}), which for the flat gauge takes the form
\be
N_{(1)}
=-\e N\zeta_1.
\ee

Putting all these steps together, we find the second-order action to be 
\bea
S_2\!=\!\!\int \!\! \d^4 x \frac{a^3\e}{\kappa^2}\!&\Bigg\{&\!\!\!
\frac{1}{c_{ad}^2}\frac{1}{N}
\dot{Z}_1^2
+\frac{1}{N}
\dot{Z}_2^2
+NH^2\Bigg[\lh 
\xi^\parallel+3\hpa+\lh\frac{2\e}{c^2_{ad}}-\Lambda_{\bar{0}}^\parallel\rh 
\tilde{\eta}^\parallel
+\frac{\e^2}{c_{ad}^2}
\rh\zeta_1^2
\nn\\
&&\!\!\!\!\!\!\!\!\!\!
+\lh2\e\hat{R}_{1221}-3\lh\chi-\frac{\dot{\e}}{2\e NH}\rh\!\!\rh\zeta_2^2
+2\lh \xi^\perp+3\hpe
-\Lambda_{\bar{0}}^\perp\tilde{\eta}^\parallel
+\e(\hpe-\Lambda_{\bar{0}}^\perp)\rh\!
\zeta_1\zeta_2\Bigg]\nn\\
&&\!\!\!\!\!\!\!\!\!\!
+2H
\dot{Z}_1
\Bigg[\lh\Lambda_{\bar{0}}^\parallel-\frac{\e}{c_{ad}^2}\rh\zeta_1+\Lambda_{\bar{0}}^\perp\zeta_2\Bigg]
-\frac{N}{a^2}\lh\lh\partial_i\zeta_1\rh^2
+c_{en}^2\lh\partial_i\zeta_2\rh^2\rh\Bigg\},
\eea
where $\hat{R}_{klmn}=R_{klmn}/(\kappa^2 \Xi)$. 
This is the second-order action governing the evolution of the first-order adiabatic and isocurvature perturbations. In order to compare with the canonical case, studied in \cite{Tzavara:2011hn}, we perform three integrations by parts and use (\ref{derzeta}) to rewrite the action in the form
\bea
S_2\!=\!\!\int \!\! \d^4 x \frac{a^3\e}{N\kappa^2}\!\!\!&\Bigg\{&\!\!\!
\frac{\dot{\zeta}_1^2}{c_{ad}^2}+\dot{\zeta}_2^2
+\lh 2\chi+
\tilde{\eta}^\parallel- 
\hpa
\rh 
NH\dot{\zeta}_2\zeta_2
-2NH\lh\hpe+(\hpe-\Lambda_{\bar{0}}^\perp) 
c_{ad}^2
\rh \frac{\dot{\zeta}_1}{c^2_{ad}}\zeta_2\\
&&\!\!\!\!\!\!\!\!
+\Bigg[2\e\hat{R}_{1221}
-3\Lambda^{221}-2\e^2
+\tilde{\eta}^\parallel
\lh 2\chi-\hpa-\Lambda^{22}_{\bar{0}}\rh+\frac{\hpe}{c_{ad}^2}\lh 
\hpe+2(\hpe-\Lambda_{\bar{0}}^\perp) c_{ad}^2\rh\nn\\
&&\!\!\!\!
+\frac{2}{3}\hpe\!\!\lh\!\xi^\perp-\tilde{\eta}^\parallel
\Lambda_{\bar{0}}^\perp\!\rh\!
-3\e\!\lh\!\!\frac{\tilde{\eta}^\parallel+\hpa}{2}-\chi\!\rh\!\!
\Bigg]\!\zeta_2^2(NH)^2
-\!\frac{N^2}{a^2}\!\lh\!\lh\partial_i\zeta_1\rh^2
+c_{en}^2\lh\partial_i\zeta_2\rh^2\!\rh\!\!\!
\Bigg\}.\nn
\eea
Note that although the term $\dot{\zeta}_2\zeta_2$ can be recast as $\zeta_2^2$ by integration by parts, we choose to keep it in this form in order to make contact with the canonical action computed in \cite{Tzavara:2011hn}, as well as to visually separate the dominant terms (the coefficients of $\dot{\zeta}_2\zeta_2$
and $\dot{\zeta}_1\zeta_2$) for the equations of motion in the long-wavelength limit, discussed in subsection \ref{long}, from the higher-order slow-roll terms in the coefficient of $\zeta_2^2$. 
As expected the time derivative of the adiabatic perturbation is always accompanied by the inverse of its sound speed. On the other hand, the entropic sound speed appears only in the spatial gradient term of the isocurvature perturbation (compare with our definition of the sound speeds (\ref{sound})). In the limit where $c_{ad}^2=c_{en}^2=1$ and  $\Lambda_{\bar{0}}^m=\Lambda_{\bar{0}}^{mn}=
\hat{R}_{1221}=0$ (and hence $\hpa=\tilde{\eta}^\parallel$) we recover the action of the canonical kinetic term models with trivial field metric. 

Interestingly enough, the first order perpendicular slow-roll parameters $\hpe$ and $\Lambda_{\bar{0}}^\perp$ also appear in a combination equivalent to the one of
$\tilde{\eta}^\parallel$. 
Although this combination does not occur naturally from the derivative of one of the background quantities (as opposed to $\tilde{\eta}^\parallel$ which is the time derivative of $X$ given by (\ref{tildedef})), we will use from now on the unifying short-hand notation
\be
\hat{\eta}^A=(\eta^A-\Lambda_{\bar{0}}^A)c_{ad}^2,
\label{hat_def}
\ee
and hence for the parallel component one can use interchangeably $\tilde{\eta}^\parallel$ or $\hat{\eta}^\parallel$.

Using the second-order action we can derive the equations of motion for the first-order adiabatic and isocurvature perturbations, which we will denote by $\delta\cL/\delta\zeta_m$, where $\cL$ is the second-order Lagrangian density:
\bea
\frac{\delta \cL}{\delta\zeta_1}&=&
-\frac{2a^3\e}{N}\Bigg\{\frac{\ddot{\zeta}_{1}}{c_{ad}^2}
+\frac{NH}{c_{ad}^2}\lh 
3+2\e+\hpa+\hat{\eta}^\parallel
-\frac{2\dot{c}_{ad}}{c_{ad}NH}
-\frac{\dot{N}}{NH^2}
\rh\dot{\zeta}_1
-\frac{NH}{c_{ad}^2}(\hpe+\hat{\eta}^\perp)\dot{\zeta}_2
\nn\\
&&\qquad\quad\!
-(NH)^2
\Bigg[ 
\lh\frac{1}{c_{ad}^2}+1\rh\lh\xi^\perp+
\hpe(3+2\e+\hat{\eta}^\parallel-\hpa)\rh
-3\Lambda_{\bar{0}}^{12}
-(3+2\e+3\hat{\eta}^\parallel)
\Lambda_{\bar{0}}^\perp
\nn\\
&&\qquad\qquad\qquad\quad
-\hat{\eta}^\parallel
\Lambda_{\bar{0}\bar{0}}^\perp
+\hpe\Lambda_{\bar{0}}^\parallel
-\frac{2\dot{c}_{ad}}{NHc_{ad}^3}\hpe 
\Bigg]\zeta_2 
\Bigg\}+2a\e N\partial^2\zeta_1
\nn\\
& 
=&\frac{\d}{\d t}\lh-2a^3\frac{\partial^2\lambda}{N}\rh
+2a\e N\partial^2\zeta_1=0,
\qquad\qquad
\partial^2\lambda=\frac{\e}{c_{ad}^2}
\lh\dot{\zeta}_1
-NH(\hpe
+\hat{\eta}^\perp
) 
\zeta_2\rh
\label{eqmo}
\eea
\bea
\frac{\delta \cL}{\delta\zeta_2}&=&
-\frac{2a^3\e}{N}\Bigg\{
\ddot{\zeta}_{2}
+NH\lh 
3+2\e+\hpa+\hat{\eta}^\parallel
-\frac{\dot{N}}{NH^2}
\rh\dot{\zeta}_2+
NH(\hpe+\hat{\eta}^\perp)\frac{\dot{\zeta}_1}{c_{ad}^2}
\nn\\
&&
\qquad\quad\!
+(NH)^2\Bigg[2\e^2+\frac{1}{2}(1+c_{ad}^2)\xi^\parallel+3\chi
+\e(3\hpa+\hat{\eta}^\parallel)
+\frac{1}{2}(-1+c_{ad}^2)\xi^\perp
\nn\\
&&
\qquad\qquad\qquad\ \ \ 
-2\e \hat{R}_{1221}
+(\hat{\eta}^\parallel+\hat{\eta}^\perp)
\frac{\dot{c}_{ad}}{c_{ad}NH}
+\frac{1}{2}\hat{\eta}^\parallel
\lh\hpa(1-2c_{ad}^2)+\hat{\eta}^\parallel
\rh
\nn\\
&&
\qquad\qquad\qquad\ \ \ 
+\frac{1}{2}\hpe(\hpa-\hat{\eta}^\parallel)
+\frac{1}{2}\lh3+2\e+\hat{\eta}^\parallel
+\hpe\lh1+\frac{2}{c_{ad}^2}\rh
\rh(\hpe-\hat{\eta}^\perp)
\nn\\
&&
\qquad\qquad\qquad\ \ \ 
+c_{ad}^2\lh\hat{\eta}^\parallel
(\Lambda_{\bar{0}\bar{0}}^\parallel
+\Lambda_{\bar{0}\bar{0}}^\perp)
+3(\Lambda_{\bar{0}}^{11}
+\Lambda_{\bar{0}}^{12})\rh
+\Lambda_{\bar{0}}^\perp
(3\hpe-c_{ad}^2\hat{\eta}^\parallel)\Bigg]
\zeta_2\Bigg\}
\nn\\
&&\qquad\quad\!
+2a\e N c_{en}^2\partial^2\zeta_2=0,
\label{eqmo2}
\eea
where $\partial^2\equiv\partial_i\partial^i$. The combination $\lambda$ is defined by the first-order energy constraint equation (\ref{en}), which in 
the flat gauge takes the form  
\be
\partial^iN^{(1)}_{i} = \partial^2\lambda.
\ee
Its importance will become clear in subsection \ref{long}. 
One can easily verify that both equations reduce 
to the related canonical expressions in \cite{Tzavara:2011hn} (where a specific choice of time coordinate was made: the number of e-folds, which means $NH=1$ and hence $\dot{N}/N=\e$), assuming $c_{ad}^2=c_{en}^2=1$ and  $\Lambda_{\bar{0}}^m=\Lambda_{\bar{0}}^{mn}=
\hat{R}_{1221}=0$ and $\hat{\eta}^A=\tilde{\eta}^A=\eta^A$ .

\section{Third-order action}\label{sec.third}

Next, we perturb the action (\ref{actionexact}) to third order, in order to find the cubic interactions of the gauge-invariant cosmological perturbations needed to calculate the related non-Gaussianity for the diagonal class of models. We work again in the flat gauge. 
After a long but straightforward computation (similar to the one in 
\cite{Tzavara:2011hn} for the canonical case) we find the final result
\bea
S_{3(1)}
&=&
\int\d^4 x \frac{a^3\e}{\kappa^2N}\Bigg\{
-\frac{1}{NH}\lh
\frac{K_1}{3}\dot{\zeta}^2_1
+ K_2\dot{\zeta}^2_2
\rh\dot{\zeta}_1
-M_1^m\zeta_m
\dot{\zeta}_1^2 
-2K_2
E^m\zeta_m
\dot{\zeta}_1\dot{\zeta}_2
-M_2^m\zeta_m\dot{\zeta}_2^2
\nn\\
&&
-NH\Biggl[\bigg(
K_2E^mE^n\zeta_m\zeta_n+\lh2M_1^m\zeta_m
+K_1
(\hat{\eta}^\parallel\zeta_1-\hpe\zeta_2)
\rh(\hat{\eta}^\parallel\zeta_1-\hpe\zeta_2)
+3\Lambda_{\bar{0}}^{mn}\zeta_m\zeta_n
\nn\\
&&\qquad\qquad 
-\frac{\e}{c_{ad}^2}\lh(2\e+\hpa)\zeta_1+2\hpe\zeta_2
\rh\zeta_1+\frac{2 \e}{3} \left(\frac{3}{c_{ad}^2}+1\right)
\hat{R}_{1221} \zeta_2 ^2
\bigg)\dot{\zeta}_1\nn\\
&&\qquad\quad
+\left(
-2 E^nM_2^m\zeta_m\zeta_n-\frac{8 \e}{3} 
\hat{R}_{1221} \zeta_1 \zeta_2 \right)\dot{\zeta}_2 \Biggr]
\nn\\
&&
-(NH)^2\Bigg\{
E^kE^mM_2^n\zeta_k\zeta_m\zeta_n+
(\hat{\eta}^\parallel\zeta_1-\hpe\zeta_2)^2
M_1^m\zeta_m
+\frac{2}{3}K_1
(\hat{\eta}^\parallel\zeta_1-\hpe\zeta_2)^3
\nn\\
&&\qquad\qquad
-3(\hat{\eta}^\parallel\zeta_1-\hpe\zeta_2)
\Lambda_{\bar{0}}^{mn}\zeta_m\zeta_n
+\left[
\e
\lh3\hpa-\hat{\eta}^\parallel\lh 
\hpa-\frac{\hat{\eta}^\parallel}{c_{ad}^2}
\rh
+\xi^\parallel\rh
+3\Lambda^{111}
\right]\zeta_1^3
\nn\\
&&\qquad\qquad
-\e\Bigg[
\hpe\lh
2\hat{\eta}^\parallel-\frac{\hpa}{c_{ad}^2}-6
\rh
-\frac{\hat{\eta}^\perp
(\hpa+2\hat{\eta}^\parallel)}{c_{ad}^2}
-\frac{2\e(\hpe+\hat{\eta}^\perp)}{c_{ad}^2}-2\xi^\perp
\nn\\
&&\qquad\qquad\qquad
+ 2 \eta^{\perp} \hat{R}_{1221} - 9 \frac{\Lambda^{211}}{\e}
\Bigg] \zeta_1^2\zeta_2
+\left[3\Lambda^{222}-\epsilon \eta^\perp \frac23 
\left( \frac{3}{c_{ad}^2}+1\right)\hat{R}_{1221}\right]\zeta_2^3
\nonumber\\
&&\qquad\qquad
-\e\Bigg[
3\lh\chi-\frac{\hpa+\hat{\eta}^\parallel}{2}-\e\rh 
+2 \lh\e-\hat{\eta}^{\parallel} (\frac{1}{c_{ad}^2}-1)
+\frac23 (\eta^{\parallel}+3) \rh \hat{R}_{1221}
\nonumber\\
&&\qquad\qquad\qquad
-\frac{2\hpe(\hpe+\hat{\eta}^\perp)}{c_{ad}^2} 
-9\frac{\Lambda^{221}}{\e}
\Bigg]\zeta_2^2\zeta_1
+\frac{2\e}{3}\left[ \frac{1}{\kappa} \sqrt{\frac{2 \e}{\Xi}}
\hat{R}^{;m} _{1221} + 4 \hat{R}_{1221} \Lambda^m _{\bar{0}}\right] \zeta^2_2\zeta_m
\Bigg\}
\nn\\
&&
+\frac{1}{a^2}\Bigg[-2\partial^i\lambda\lh\frac{\dot{\zeta}_1}{c_{ad}^2}\partial_i\zeta_1+\dot{\zeta}_2
\partial_i\zeta_2\rh 
\nn\\
&&\qquad
+(c_{en}^2-1)\Bigg(
N^2\lh(\e+\hat{\eta}^\parallel)\zeta_2^2
\partial^2
\zeta_1+\hpe\zeta_1^2\partial^2\zeta_2\rh 
-\frac{2N}{H}\partial^i\zeta_2\partial_i\zeta_1
\dot{\zeta}_2
\Bigg)
\nn\\
&&\qquad
+2NH\frac{\hpe+\hat{\eta}^\perp}{c_{ad}^2}\zeta_2\partial_i\lambda
\partial^i\zeta_1
-2NH(\e+\hat{\eta}^\parallel)\zeta_2
\partial_i\lambda
\partial^i\zeta_2
+\frac{1}{2}\zeta_1\lh(\partial_i\partial_j\lambda)^2
-(\partial^2\lambda)^2
\rh
\nn\\
&&\qquad
+(\partial\zeta_1)^2\Bigg(
N^2\lh(\e+\hpa-\hat{\eta}^\parallel)
\zeta_1+\frac{(2c_{ad}^2-1)\hpe
-\hat{\eta}^\perp}{c_{ad}^2}\zeta_2\rh
+\frac{N}{H}\lh\frac{1}{c_{ad}^2}-1\rh\dot{\zeta}_1\Bigg)
\nn\\
&&\qquad
+(\partial\zeta_2)^2\Bigg(
N^2\lh
(c_{en}^2\e+\Lambda_{\bar{2}}^\parallel+
\Lambda_{\bar{2}\bar{0}}
\hat{\eta}^\parallel)
\zeta_1
+(\Lambda_{\bar{2}}^\perp-
\Lambda_{\bar{2}\bar{0}}\hpe)
\zeta_2\rh
+\frac{N}{H}\Lambda_{\bar{2}\bar{0}}\dot{\zeta}_1\Bigg)
\Bigg]
\Bigg\}.
\label{third_order_action_gen}
\eea
Here we defined the combinations 
\bea
\label{def_k_m_e}
&&K_1=
\Lambda_{\bar{0}\bar{0}\bar{0}}
+3\lh\frac{1}{c_{ad}^2}-1\rh,\quad 
K_2=2\lh1-c_{en}^2\rh+\Lambda_{\bar{1}\bar{1}\bar{0}}
+\Lambda_{\bar{2}\bar{0}},\quad
E^A=\hpe e_1^A
+(\e+\tilde{\eta}^\parallel)e_2^A,\\ 
&&M_1^A=
\Lambda_{\bar{0}\bar{0}}^A
+\Lambda_{\bar{0}}^A+K_1
\tilde{\eta}^A 
-\frac{\e}{c_{ad}^2}e_1^A- 2K_1\hpe e_2^A,\quad 
M_2^A=\Lambda_{\bar{1}\bar{1}}^A
+\Lambda_{\bar{2}}^A
+K_2
\tilde{\eta}^A-\e e_1^A-2K_2\hpe e_2^A.\nn
\eea
Notice that in the above definitions (\ref{def_k_m_e}) 
only $\tilde{\eta}^A$ occurs, not $\hat{\eta}^A$. On the contrary, in the 
action (\ref{third_order_action_gen}) only $\hat{\eta}^A$ occurs explicitly.
We remind the reader that $\tilde{\eta}^A$  and $\hat{\eta}^A$
are defined respectively in (\ref{tildedef}) and (\ref{hat_def})
and that $\tilde{\eta}^\perp=\hpe$, while  
$\hat{\eta}^\parallel=\tilde{\eta}^\parallel$.
 
As was explained in \cite{Tzavara:2011hn}, the cubic part of the action 
$S_{3(1)}$ is not gauge-invariant on its own, one needs to add the contributions
of the second-order fields, $S_{3(2)}$, which we will do next.
The canonical case study in \cite{Tzavara:2011hn} taught us that these extra terms can be rewritten after a few redefinitions as quadratic terms of the gauge-invariant perturbations, multiplying the second-order equations of motion $\delta\cL/\delta\zeta_m$. Performing a redefinition of the perturbations these can finally be removed, since they cancel with the new contributions arising from the second-order action. However, this redefinition is important since it contributes to in principle observable quantities and notably to $\f^{(4)}$, the parameter of non-Gaussianity due to products of two power spectra. See \cite{Tzavara:2011hn} for details and the end of subsection \ref{long} for an example.

The aforementioned redefinition accounts for the extra quadratic terms of first-order perturbations, needed to construct the second-order gauge-invariant perturbations from the second-order non gauge-invariant fields. Hence, one can calculate them either by performing integrations by parts in the action or by finding the second-order gauge transformation. The former calculation being a rather complicated procedure, we choose to find the second-order gauge transformation of the energy density perturbation and rewrite it in terms of the field perturbations as was done in \cite{Tzavara:2011hn}. In the rest of this section we will focus on these terms needed to complete the third-order study. 

The first and second-order gauge transformations for any scalar quantity $A$ read
\bea
&&\tilde{A}_{(1)}=A_{(1)}
+\dot{A}T_{(1)}\nn\\  
&&\tilde{A}_{(2)}=A_{(2)}
+\dot{A}T_{(2)}+T_{(1)}\lh
2\dot{A}_{(1)}+\dot{A}\dot{T}_{(1)}
+\ddot{A}T_{(1)}\rh,\label{gauge2}
\eea 
where $T_{(r)}$ is the change of the time coordinate when doing a gauge transformation to the tilded gauge. More details on gauge transformations can be found in appendix \ref{app2}.  
Notice that in order to lighten the notation we shall not include the $(1)$ subscripts for the first-order perturbations in the rest of the paper. 
Applying the above transformation to the energy density perturbation at second order and setting the tilded gauge to be the uniform energy density gauge $\tilde{\rho}_{(r)}=0$ we can find the second-order time shift $T_{(2)}$. Rewriting it in terms of the first-order gauge-invariant quantities, it takes the form 
\be
T_{(2)}=-\frac{\rho_{(2)}}{\dot{\rho}}
+\frac{\zeta_1-\alpha}{NH}\lh
\frac{\dot{\zeta}_1-\dot{\alpha}}{NH}
+\lh\zeta_1-\alpha\rh\lh\hpa+\tilde{\eta}
^\parallel\rh 
\rh.  
\ee
This second-order time shift is needed to construct the second-order gauge-invariant adiabatic perturbation, which in the uniform energy density gauge coincides with the curvature perturbation:
\be
\frac{1}{2}\zeta_{1(2)}\equiv\frac{1}{2}\tilde{\alpha}_{(2)}=\frac{1}{2}\alpha_{(2)}+
\frac{1}{2}T_{(2)}NH+\frac{1}{2}
\lh\zeta_1-\alpha\rh 
\frac{\dot{\zeta}_1+\dot{\alpha}}{NH}+\partial^2\cB. 
\ee
In the above equation the $\partial^2\cB$ terms account for spatial gradient terms that vanish outside the horizon.  Their explicit form is given in appendix \ref{app2}.   We want to rewrite $T_{(2)}$ in terms of the field perturbations, since these are more appropriate during an inflationary period. To that end, one needs to use the second-order $0i$ Einstein equation, 
\bea
&&-\e NH\Big[
Q_{1(2)}+NH\frac{\rho_{(2)}}{\dot{\rho}}
+(\e-\hpa)\lh\zeta_{1}
-\alpha\rh^2
-(\e+\tilde{\eta}^\parallel)\zeta_{2}^2
-2\hpe\zeta_{2}\lh\zeta_{1}
-\alpha\rh 
\nn\\&&\qquad\quad
-\frac{2}{NH}\partial^{-2}\partial^i
\lh
\dot{\zeta}_{2}
\partial_i\zeta_{2}\rh \Big]
+\partial^2\cA
=0,
\label{r2n}
\eea
where $\partial^2\cA$ are terms involving spatial gradients and hence they can be ignored on super-horizon scales. Their form can be found in appendix \ref{app2}.   We also defined the short-hand notation
\be 
Q_{m(r)}=-\frac{H}{\tilde{\Pi}}e_{mA}\varphi^A_{(r)}.
\ee 
Putting everything together we find
\bea
&&\frac{1}{2}\zeta_{1(2)}=\frac{1}{2}\alpha_{(2)}+
\frac{1}{2}Q_{1(2)}+f_1,\nn\\
&&f_1=
\frac{\e+\tilde{\eta}^\parallel}{2}\Big[\lh\zeta_1
-\alpha\rh^2
-\zeta_{2}^2\Big]
+\lh\zeta_1-\alpha\rh
\lh  
\frac{\dot{\zeta}_1}{NH}
-\hpe\zeta_{2}\rh 
-\frac{1}{NH}\partial^{-2}\partial^i
\lh
\dot{\zeta}_{2}
\partial_i\zeta_{2}\rh\nn\\
&&\qquad -\frac{1}{2NH\e}\partial^2\cA+\partial^2\cB.
\label{redad}
\eea
This is the gauge transformation of the second-order adiabatic perturbation for an arbitrary gauge. In the flat gauge, where $\alpha=0$, one recovers the gauge transformation found in \cite{Tzavara:2011hn}, taking into account that in the canonical case $\tilde{\eta}^\parallel=\hpa$. Notice that the only evidence of the non-canonical kinetic terms in the above expression is actually $\tilde{\eta}^\parallel$, since the definition of the gauge-invariant perturbations depends only on (\ref{gauge2}) and the gauge transformation of $\alpha$ (see appendix~\ref{app2}). 

We can repeat this procedure for the second-order  isocurvature perturbation. After performing the second-order gauge transformation for the relevant field quantity $Q_{2(r)}$, we get
\be 
\frac{1}{2}\tilde{Q}_{2(2)}=\frac{1}{2}Q_{2(2)}+
\lh\zeta_{1}-\alpha\rh\Big[
(\e+\tilde{\eta}^\parallel)\zeta_{2}+\frac{1}{HN}\dot{\zeta}_{2}+\frac{\hpe}{2}\lh\zeta_{1}-\alpha\rh 
\Big].\label{Q2tr}
\ee
However, this quantity does not correspond to our isocurvature perturbation at second order (it is not orthogonal to the adiabatic perturbation). It turns out that one needs to consider the gradient of the fully non-linear quantity in order to properly define the correct combination (for details see \cite{Tzavara:2011hn}):
\be 
\frac{1}{2}\zeta_{2(2)}\equiv\frac{1}{2}\tilde{Q}_{2(2)}
-\frac{\hpe}{2}\zeta_{2}^2
+\frac{1}{NH}\partial^{-2}\partial^j\lh\dot{\zeta}_{1}
\partial_j\zeta_{2}\rh 
\ee
and hence find the second-order gauge transformation for the isocurvature perturbation
\bea
&&
\frac{1}{2}\zeta_{2(2)}
=\frac{1}{2}Q_{2(2)}+f_2,\\
&&f_2=
\frac{\hpe}{2}\Big[\lh\zeta_{1}-\alpha\rh^2
-\zeta_{2}^2\Big]
+(\e+\tilde{\eta}^\parallel)\lh\zeta_{1}-\alpha\rh \zeta_{2}
+\frac{1}{NH}\dot{\zeta}_{2}\lh
\zeta_1-\alpha\rh 
+\frac{1}{NH}\partial^{-2}\partial^j\lh\dot{\zeta}_{1}
\partial_j\zeta_{2}\rh.\nn
\eea
Setting $\alpha=0$ and $\tilde{\eta}^\parallel=\hpa$ one recovers the flat gauge expression for the isocurvature perturbation found in \cite{Tzavara:2011hn}.

Having found the gauge transformations for the cosmological perturbations, the second-order field contributions in the action take the form
\be 
S_{(3)2}=\int\d^4x\frac{\delta\cL_2}{\delta\zeta_m}f_m|_{\alpha=0},
\ee
where $\delta\cL/\delta\zeta_m$ are the equations of motion for the first-order perturbations (\ref{eqmo}--\ref{eqmo2}). 
The total third-order action, which is gauge-invariant, is then $S_{3(1)} + S_{3(2)}$ with 
$S_{3(1)}$ given in (\ref{third_order_action_gen}).  
One can compute the relevant redefinition of the adiabatic and isocurvature perturbation in the action as (see \cite{Tzavara:2011hn})
\be 
\zeta_m+\frac{1}{2}\zeta_{m(2)}=\zeta_{mc}
+f_m\Big|_{\alpha=0},\label{rule}
\ee
where $\zeta_{mc}$ are the redefined perturbations. 

With this discussion, we close the study of the third-order action and second-order perturbations. We have found the cubic interactions of the gauge-invariant perturbations, as well as the quadratic redefinition of the perturbations. With these tools in hand, one can compute the related second-order measurable quantities using the in-in formalism, i.e.\ the third-order correlation function for the adiabatic and isocurvature perturbations. We leave these computations for a future paper.

\section{Applications}\label{sec.applications}

In this section, we will illustrate our general results with some concrete examples. In particular, we will study the consequences of our results  in the long-wavelength limit, i.e.\ outside the cosmological horizon,  as well as for the symmetrized and well-studied case of DBI inflation. 

\subsection{Long-wavelength limit}\label{long}

We shall first consider the long-wavelength limit. By this term we mean keeping only zeroth order terms in a spatial gradient expansion and hence ignoring any  spatial gradient terms. This assumption follows logically from the fact that outside the horizon the wavelength of the perturbation $k$, related to gradient terms, is much smaller than the Hubble length scale $H$, related to time derivative terms. 

Applying this hypothesis at second order, we find for  the equation of motion for the adiabatic perturbation
\be 
\dot{\zeta}_1
=NH(\hpe+\hat{\eta}^\perp)
\zeta_2
\label{hc1}
\ee
and hence outside the horizon the adiabatic perturbation is only sourced by the isocurvature one, as in the canonical case. However, the sourcing also depends  on the adiabatic sound speed, as well as on the contribution of $\Lambda_{\bar{0}}^\perp$, given that $\hat{\eta}^\perp = (\eta^\perp-\Lambda_{\bar{0}}^\perp)c_{ad}^2$. 
In the limit $c_{ad}^2\rightarrow 0$ the sourcing of the adiabatic perturbation by the isocurvature one is only half of what it is in the canonical case.
Notice that in the super-horizon regime, as in the canonical case, the equation for the adiabatic mode (\ref{eqmo}) coincides with the first-order momentum constraint equation
\be
\partial^2\lambda=0. 
\ee

For the isocurvature equation one needs to assume slow roll in addition to the long-wavelength approximation in order to further simplify the evolution equation. By the slow-roll assumption we also mean here that $\dot{c}_{ad}\ll NHc_{ad}$. Applying these approximations to (\ref{eqmo2}) we find
\be
\dot{\zeta}_2
=\lh\frac{2}{3}\e\hat{R}_{1221}-\chi-\frac{
\tilde{\eta}^\parallel-\hpa}{2}\rh NH\zeta_2.
\label{hc2}
\ee
As in the canonical case the isocurvature component evolves independently from the adiabatic one. Its evolution depends on the generalized slow-roll parameters
and the adiabatic sound speed, since $\tilde{\eta}^\parallel = (\eta^\parallel - \Lambda_{\bar{0}}^\parallel)c_{ad}^2$, as well as on the curvature of the field space $\hat{R}_{1221}$.   On the contrary, the adiabatic perturbation is not affected by the field curvature. Notice finally that the entropic sound speed does not appear in these equations and hence plays no role in the super-horizon evolution of the perturbations. 

Given a specific model of inflation, these equations can  be perturbed to second order and in principle can be solved using Green's functions within the long-wavelength formalism. This would allow not only for the study of the super-horizon evolution of the perturbations, but also of the scale-dependence of the related super-horizon non-Gaussianity. Since this is beyond the scope of this paper, we only give here the super-horizon parameter of non-Gaussianity directly after horizon crossing, $f_{\mathrm{NL}*}^{(4)}$. It can be found from the redefinition of the adiabatic perturbation (\ref{redad}) and the rule (\ref{rule}). This redefinition will contribute to the horizon-crossing third-order correlation function as
\be
\langle\zeta_{1(2)}\zeta_1\zeta_1\rangle=(\e_*+\tilde{\eta}^\parallel_*)\langle\zeta_1\zeta_1\rangle\langle\zeta_1\zeta_1
\rangle,
\ee
where we used the slow-roll result that right outside the horizon $\langle\zeta_1\zeta_2\rangle=0$.
Hence one can find for $\f^{(4)}$ at horizon crossing
\be
-\frac{6}{5}f^{(4)}_{\mathrm{NL}*}=\e_*+\tilde{\eta}^\parallel_*
=\e_*+c_{ad}^2\lh\hpas-
\Lambda_{\bar{0}*}^\parallel\rh. 
\ee
The star indicates that quantities are evaluated at horizon crossing.

\subsection{DBI model}\label{subsec_dbi}

Here we consider DBI inflation \cite{Silverstein:2003hf,Alishahiha:2004eh},
which is motivated by string theory and known to be able to give
large equilateral-type non-Gaussianity.
In DBI inflation, the inflaton is identified with the position of a moving
D3 brane whose dynamics are described by the DBI action. 
Since the position of the brane in each compact direction is described by
a scalar field, DBI inflation is naturally a multiple-field model 
\cite{Easson:2007dh}.
In the two-field DBI model, the Lagrangian is given by
\begin{eqnarray}
\bar{P}(\bar{Y}, \varphi) &=& 
-\frac{1}{\bar{f}(\varphi)}
\left(\sqrt{1-2 \bar{f}(\varphi) \bar{Y}} -1\right)-W(\varphi)\,,\nonumber\\
&&\;\;\;\;\bar{Y} \equiv 
\bar{X} - \bar{f}(\varphi) \left(\bar{X}^2 - \bar{S}\right)\,,
\;\;\;\;
\bar{S} \equiv G_{AC} G_{BD} \bar{X}^{AB} \bar{X}^{CD}\,,
\label{DBI_model}
\end{eqnarray}
where $W$ is the potential and $f$ is a function of the scalar fields
that reflects the information of the geometry of the compactified internal space.
Here, we do not need to specify the form of these functions, our results
are valid for arbitrary $W$ and $f$.
It is worth mentioning that 
$\bar{Y}$ is equal to $X$ when evaluated at the background level, which we write as
\begin{eqnarray}
\bar{Y} \rightarrow X\,.
\end{eqnarray}

In this model, the variable $\bar{\gamma}$ defined by
\begin{eqnarray}
\bar{\gamma} \equiv \sqrt{1-2 \bar{f}(\varphi) \bar{Y}}\,,
\end{eqnarray}
with the background behaviour
\begin{eqnarray}
\bar{\gamma} \rightarrow   
\gamma \equiv \sqrt{1-2 f(\phi) X} \,,
\end{eqnarray}
plays an important role. 
Actually, based on the definition of the sound speed given 
in Eq.~(\ref{sound_speed_diagonal}), we can show that
\begin{eqnarray}
\gamma = c_{ad} = c_{en} \equiv c_s\,.
\end{eqnarray}
In general, $\Lambda$ parameters with a different number of kinetic
derivatives with respect to $X^{AB}$ are not related through a common function. 
But as we will show, in the DBI model, these 
$\Lambda$ parameters are related through $c_s$ as long as 
the number of covariant field derivatives is the same.
Then in addition to the field curvature, which always appears
when we consider non-flat field spaces, the new physical ingredients
in DBI inflation compared to a canonical two-field inflationary model
are only the sound speed and the derivatives
of the function $f$.

We briefly summarize the quantities appearing in the action
in the DBI model. For the intermediate steps to obtain these,
see Appendix~\ref{dbi_intermediate}.
The $\epsilon$ parameter is given by
\begin{eqnarray}
\epsilon =  c_s\frac{\kappa^2 \Pi^2}{2H^2}\,.
\end{eqnarray}
For the linear terms in the expansion of $P$, we find that
\begin{eqnarray}
P_{\bar{0}} &=& \frac{1}{c_s} = \Xi  \,,\quad 
P_{\bar{2}} =c_s^2 P_{\bar{0}}\,,\quad
\eta^A = -3 e^A _1 + \frac{1}{H \Pi} 
\left(\frac{(1-c_s)^2}{2 f^2 c_s} f^{;A} - W^{;A}\right)\,,
\end{eqnarray}
while $P_{\bar{1}} = 0$ as expected (required for the adiabatic and entropic 
sound speeds to be well-defined).
For the quadratic terms in the expansion of $P$, we find that
\begin{eqnarray}
&&\Lambda_{\bar{0}\bar{0}} = \frac{1-c_s^2}{c_s^2}\,,\quad
\Lambda_{\bar{1}\bar{1}} = c_s^2 \Lambda_{\bar{0}\bar{0}}\,,\quad
\Lambda_{\bar{2}\bar{0}} = -c_s^2 \Lambda_{\bar{0}\bar{0}}\,,\quad
\Lambda_{\bar{2}\bar{2}} =c_s^4 \Lambda_{\bar{0}\bar{0}}\,,\\
&&\Lambda_{\bar{0}} ^{A} = \frac{f^{;A}(1-c_s^2)^2}{2 f^2 H \Pi c_s^3}\,,
\quad \Lambda_{\bar{2}} ^{A} = -c_s^2 \Lambda_{\bar{0}} ^{A}\,,\quad
\Lambda^{AB} = \frac{c_s}{3 H^2} P^{;AB}\,,
\end{eqnarray}
while $\Lambda_{\bar{1}\bar{0}} = \Lambda_{\bar{2}\bar{1}} 
= \Lambda_{\bar{1}} ^{A}= 0$.
The $\hat{\eta}^{A}$ parameter is given by
\begin{eqnarray}
\hat{\eta}^{A} = c_s^2 \left[-3 e^A _1 -\frac{1}{H \Pi}
\left(\frac{(1-c_s)^2 (1+2 c_s)}{2 f^2 c_s^3} f^{;A} + W^{; A}\right)\right]\,.
\end{eqnarray}
Notice that while $\hat{\eta}^{\parallel} = \tilde{\eta}^{\parallel}$,
$\hat{\eta}^{\perp} \neq \tilde{\eta}^{\perp} (= \eta^{\perp}) $.
Similarly, for the cubic terms in the expansion of $P$, we find that
\begin{eqnarray}
\Lambda_{\bar{0}\bar{0}\bar{0}} &=& \frac{3(1-c_s^2)^2}{c_s^4}\,,
\quad
\Lambda_{\bar{1}\bar{1}\bar{0}} = \frac{c_s^2}{3}
\Lambda_{\bar{0}\bar{0}\bar{0}}\,,\quad
\Lambda_{\bar{2}\bar{0}\bar{0}} = -\frac{c_s^2}{3}
\Lambda_{\bar{0}\bar{0}\bar{0}}\,,\\
\Lambda_{\bar{2}\bar{2}\bar{0}} &=& -\frac{c_s^4}{3}
\Lambda_{\bar{0}\bar{0}\bar{0}}\,,\quad
\Lambda_{\bar{2}\bar{1}\bar{1}} = \frac{c_s^4}{3}
\Lambda_{\bar{0}\bar{0}\bar{0}}\,,\quad
\Lambda_{\bar{2}\bar{2}\bar{2}} = c_s^6
\Lambda_{\bar{0}\bar{0}\bar{0}}\,,\\
\Lambda_{\bar{0}\bar{0}} ^A &=& \left(\frac{3}{c_s^2}-1\right)\Lambda_{\bar{0}} ^A
\,,\quad
\Lambda_{\bar{1}\bar{1}} ^A = (1+c_s^2)\Lambda_{\bar{0}} ^A\,,\quad
\Lambda_{\bar{2}\bar{0}} ^A = -(1+c_s^2)\Lambda_{\bar{0}} ^A\,,\\
\Lambda_{\bar{2}\bar{2}} ^A &=& -c_s^2 (1-3c_s^2)\Lambda_{\bar{0}} ^A\,,\quad
\Lambda_{\bar{0}} ^{AB} = \frac{(1-c_s^2)^2}{12 f^3 H^2 c_s^4}
\left[2 f c_s^2 f^{;AB} + 3 (1-c_s^2) f^{;A} f^{;B}\right]\,,\\
\Lambda_{\bar{2}} ^{AB} &=& -\frac{(1-c_s^2)^2}{12 f^3 H^2 c_s^2}
\left[2 f c_s^2 f^{;AB} + (1-c_s^2) f^{;A} f^{;B}\right]\,,\quad
\Lambda^{ABC} = \frac{1-c_s^2}{9 f H^3 \Pi} P^{;ABC}\,,
\end{eqnarray}
while $\Lambda_{\bar{1}\bar{0}\bar{0}}= \Lambda_{\bar{1}\bar{1}\bar{1}}
=\Lambda_{\bar{2}\bar{1}\bar{0}}=\Lambda_{\bar{2}\bar{2}\bar{1}}
= \Lambda_{\bar{1}\bar{0}} ^A = \Lambda_{\bar{2}\bar{1}} ^A = 
\Lambda_{\bar{1}} ^{AB}=0$.
The quantities $K_1$, $K_2$, $M_1 ^A$, $M_2 ^A$ defined in (\ref{def_k_m_e})
are simplified as
\begin{eqnarray}
&&K_1 =\frac{3}{c_s^2}K_2= \frac{3(1-c_s^2)}{c_s^4}\,,\nonumber\\
&&M_1 ^A =\frac{3}{c_s^2}M_2^A+\frac{2}{c_s^2}\e e_1^A=\frac{3}{c_s^2}(\eta^A-\hat{\eta}^A)
-\frac{1}{c_s^2}\e e_1^A
+\frac{3}{c_s^2}\lh1-\frac{1}{c_s^2}\rh(\eta^\perp+
\hat{\eta}^\perp)e_2^A
\,.
\end{eqnarray}
Finally, the time derivative of $c_s$ is given by
\begin{eqnarray}
\frac{\dot{c}_s}{NH c_s} &=& \tilde{\eta}^{\parallel} - {\eta}^{\parallel}\,,
\end{eqnarray}
where we can see that  for the canonical case with $c_s=1$,
when $\tilde{\eta}^{\parallel} = {\eta}^{\parallel}$,
it becomes trivial. 
Although we do not give the explicit form of  $\xi^A$ and $\chi$
as they are long and not so interesting, they can also be expressed in terms of
the $\Lambda$ parameters above.

Inserting all the above expressions into the general third-order action
(\ref{third_order_action_gen}), we find the complete action for
the DBI model. An alternative expression is given in (\ref{DBI_third_exact}).
As a consistency check, if we take the small sound speed limit $c_s \ll 1$
together with keeping only the leading-order quantities in the slow-roll approximation, we obtain
\bea
S_{3(1)}  &=& \int d^4 x \frac{a^3 \e}{\kappa^2 N}\Bigg\{
-\frac{\dot{\zeta}_1}{NH c_s^2}
\Bigg[\frac{\dot{\zeta}_1^2}{c_{s}^2}  
+\dot{\zeta}_2^2 \Bigg] 
\nn\\&&
+\frac{N}{ H} \Bigg[ \frac{1}{c_{s}^2} 
\dot{\zeta}_1 h^{ij} \partial_i \zeta_1 \partial_j \zeta_1 
-\dot{\zeta}_1
h^{ij} \partial_i \zeta_2 \partial_j \zeta_2
+2 \dot{\zeta}_2 h^{ij} \partial_i \zeta_1 \partial_j \zeta_2\Bigg]\Bigg\},
\label{DBI_third_leading}
\eea
which was first obtained in \cite{Langlois:2008wt}.
Here it is also assumed that $\hat{R}_{1221}$ is not too large, so that when 
multiplied with a slow-roll parameter it can be neglected.
Currently, the study of the non-Gaussianity in multiple-field DBI inflation is 
limited to the case where the above conditions are satisfied 
(for example, see \cite{Kidani:2012jp}). In that case, 
based on (\ref{DBI_third_leading}) and the $\delta N$ formalism,
the amplitude of the equilateral-type bispectrum is given by \cite{Langlois:2008qf}
 \bea
f^{(3)} _{\rm NL} = -\frac{35}{105 c_s^2} \frac{1}{1+T_{\sigma s}^2}\,,
\eea
where $T_{\sigma s}$ denotes the transfer between adiabatic and entropic modes
after horizon crossing. However, observational results by Planck
show that $f^{(3)} _{\rm NL}$ is small, which suggests that
$c_s$ cannot be much smaller than one, hence invalidating the assumptions
of (\ref{DBI_third_leading}), unless $T_{\sigma s}$ is sufficiently large.
On the other hand, the action (\ref{third_order_action_gen}) or 
(\ref{DBI_third_exact}) we obtained here 
can be used directly with the in-in formalism  in order to calculate
the exact non-Gaussianity in more general set-ups: 
beyond the slow-roll or small sound speed limit.
We leave that computation for a future paper.

\section{Conclusions}\label{sec.conc}

In this paper we addressed several issues concerning generalized models of two-field inflation. Such models are for example related to extra-dimensional physical theories, after these have been compactified and hence new scalar degrees of freedom have appeared. The scalar degrees of freedom usually come along with non-trivial kinetic terms and field metrics.  The related literature is very rich since such models  offer specific signatures that can distinguish them from the canonical models, especially concerning their predictions about non-Gaussianity. 

Generalized kinetic terms induce a new type of derivative, namely the derivative of the Lagrangian with respect to the kinetic terms $X^{AB}$, while non-trivial field metrics introduce the necessity to consider covariant field derivatives respecting parallel displacement of vectors in the curved field space.  
We first considered the background properties of such models. It is clear that when studying the dynamics and deriving equations of motion, mixed covariant field and kinetic-term derivatives appear. Hence the issue of whether these derivatives commute becomes important. It turns out that they do, contrary to what was assumed in the literature so far. One only needs to take into account the appropriate  covariant relation $X_{AB;C}=0$ for the generalized coordinates, which is natural in the context of a Lagrangian formulation in curved field space.

However, the next problem one has to confront is whether one should consider the generalized momentum $\Pi^A$ of the fields or their velocity $\tilde{\Pi}^A$ to define the field-space basis and hence the parallel and perpendicular components of quantities that live in field space. We dealt with this problem by considering a restricted but still very general class of models, the diagonal class of models, where the two vectors are parallel. This is equivalent to saying that we considered models for which the kinetic terms are only contracted with tensors that are diagonal in the adiabatic-entropic basis (for example, this would exclude terms of the form $\varphi_A\varphi_BX^{AB}$ in a model with a turning field
trajectory). As a consequence, as we showed, this class of models is also 
defined by having a well-defined adiabatic and entropic sound speed.

We next built a new hierarchy of slow-roll parameters $\Lambda$ (\ref{newl}) that involve mixed derivatives of fields and kinetic terms. The pure kinetic-term derivatives are related to sound speeds, naturally appearing in all models with higher-order kinetic terms. Pure field and mixed derivatives of the Lagrangian, on the other hand, are effectively dynamical slow-roll parameters that characterize the inflationary background. We emphasize that while we call these quantities slow-roll parameters, we do not use any slow-roll approximation in this paper, all expressions are completely exact and the parameters can be large.

We derived the exact second and third-order action of the gauge-invariant adiabatic and isocurvature perturbations in terms of the covariant slow-roll parameters for the diagonal class of models. 
At third order, we not only computed the cubic action of the first-order fields,
we also considered the second-order contributions of the fields, 
which leads to a redefinition of the perturbations that 
contributes to the three-point correlation function. From this redefinition we computed the horizon-crossing part of $\f^{(4)}$, the  $\f$ parameter related to products of two power spectra. 

As an illustration, we considered the DBI model, a highly symmetrised model, in order to better understand  the different quantities involved in the theory. For the DBI model the adiabatic and entropic sound speeds coincide, and there are only two scalar functions of the fields: their potential and the function $f$, a remnant of the geometry of the internal compactified space. 
It turns out that the new physical ingredients, as compared to a canonical two-field inflationary model, are the field curvature, the sound speed and the derivatives of the function $f$.

The covariant third-order action we have computed, beyond slow-roll or super-horizon approximations,  will likely prove very useful for future calculations. As it is the exact action in terms of the gauge-invariant perturbations (as opposed to the fields) one does not need to use the $\delta N$ formalism, and hence assume slow-roll at horizon crossing, in order to calculate the related non-Gaussianity. We leave the explicit computation of the bispectrum and its $\f$ parameters for a future paper.

\section*{Acknowledgements}
S.M. was financially supported by the IN2P3 and the Labex P2IO in Orsay. 
We thank the anonymous referee for useful comments.

\appendix

\section{Gauge issues}
\label{app2}

In this appendix we present some intermediate steps of the calculation of the 
gauge transformations used in section \ref{sec.third} and of the perturbed 
$0i$ Einstein equation (\ref{r2n}).

\subsection{Gauge transformations}

In the context of perturbation theory around an homogeneous background any quantity $\bar{A}$ that has a non-zero background value can be decomposed into an homogeneous part $A(t)$ and an infinite series of perturbations as
\be
\bar{A}(t,\vc{x})=A(t)+A_{(1)}(t,\vc{x})+\frac{1}{2}A_{(2)}(t,\vc{x})+\cdots,
\ee 
where the subscripts in the parentheses denote the order of the perturbation. 

We decompose the spatial part of the ADM metric as 
\be
\bh_{ij}=a(t)^2\exp\big[2\alpha\delta_{ij}+2\chi_{ij}
\big], 
\ee
where $\chi_{ij}$ is traceless and $\alpha$ is the scalar curvature perturbation. 
As any tensor, the spatial metric can be split into a scalar, a vector and a tensor part as
\be
\chi_{ij}=D_{ij}F+F_{(i|j)}+\gamma_{ij},
\ee
where $|$ or $D_i$ is the covariant derivative on the space hypersurface, i.e.\ relative to the  spatial metric $\bh_{ij}$ (being just $\partial_i$ when acting on space scalars) and indices enclosed in parentheses are to be symmetrized. We have also introduced the operator
\be
D_{ij}\equiv D_iD_j-\frac{1}{3}\delta_{ij}D_kD^k. 
\ee 
The vector $F_i$ is divergence-free, while the pure  tensor part $\gamma_{ij}$, that represents gravity waves, is transverse, i.e.\ traceless and divergence-free:
\be 
F^i_{|i}=0,\quad 
\gamma^k_{i|k}=0
\quad\mathrm{and}\quad\gamma_i^i=0.
\ee
Finally, the shift function $N^i$ can be split up again in a scalar part and a divergence-free vector:
\be N^i=\partial^i\psi+N^i_{\perp} \quad\mathrm{with}
\quad N^i_{\perp|i}=0.
\ee

Next we will consider arbitrary coordinate transformations that up to second order take the form  \cite{Bruni:1996im,Malik:2003mv}
\be
\widetilde{x}^{\mu}=x^{\mu}+\beta_{(1)}^{\mu}
+\frac{1}{2}\lh\beta_{(1)|\nu}^{\mu}\beta^{\nu}_{(1)}+\beta_{(2)}^{\mu}\rh\qquad\mathrm{with}\qquad
\beta^0=T
\qquad\mathrm{and}\qquad 
\beta^i=\partial^i\beta+\beta^i_\perp,
\ee
where again we split the space part of the coordinate transformation in a derivative of a scalar and a divergence-free vector.
Note that the tilde in this appendix corresponds to an arbitrary gauge, it is
not necessarily the uniform energy density gauge as in section~\ref{sec.third}.
Under such coordinate transformations the perturbations of a tensor transform as \cite{Bruni:1996im}
\bea
&&\widetilde{A}_{(1)}=A_{(1)}+L_{\beta_{(1)}}A,\nn\\
&&\widetilde{A}_{(2)}=A_{(2)}+L_{\beta_{(2)}}A+
L_{\beta_{(1)}}^2A+2L_{\beta_{(1)}}A_{(1)},\label{trans}
\eea
where $L_{\beta}$ is the Lie derivative along the vector $\beta$
\be
\lh L_{\beta}A\rh^{\mu_1\mu_2...}_{\nu_1\nu_2...}=\beta^\kappa\partial_\kappa A^{\mu_1\mu_2...}_{\nu_1\nu_2...}
-\partial_\kappa\beta^{\mu_1}
A^{\kappa\mu_2...}_{\nu_1\nu_2...}-\dots +\partial_{\nu_1}\beta^{\kappa}A^{\mu_1\mu_2...}_{\kappa\nu_2...}+\dots .
\ee

Using the gauge transformation (\ref{trans}) for $g_{ij}$ one can find to second order
\bea
\tilde{\alpha}_{(2)}\delta_{ij}+\tilde{\chi}_{ij(2)}
& = & \alpha_{(2)}\delta_{ij}+\chi_{ij(2)}
+NHT_{(2)}\delta_{ij}+\beta_{2(i,j)}
+\partial_{(i}TN_{j)}  
+\partial_{(i}T\tilde{N}_{j)} 
\nn\\
&&+T\Big[\lh\dot{\alpha}+\dot{\tilde{\alpha}}
\rh\delta_{ij}
 +\dot{\chi}_{ij}+\dot{\tilde{\chi}}_{ij}\Big] 
+\beta^k\partial_k\Big[\lh \alpha+\tilde{\alpha}\rh\delta_{ij} 
  +\chi_{ij}+\tilde{\chi}_{ij}
 \Big]\nn\\
&&
+\frac{1}{2}\lh \beta^k_{,i}\beta_{k,j}-\beta_{i,k}\beta_j^{,k}\rh 
+\chi_{kj}\lh\beta^K_{,i}-\beta_i^{,k}  \rh 
+\chi_{ki}\lh\beta^k_{,j}-\beta_j^{,k}  \rh, 
\eea
where we have suppressed the $(1)$ index for first-order quantities, in order to lighten the notation. 
Taking the trace of the above equation we find
\bea
\tilde{\alpha}_{(2)}=\alpha_{(2)}+NHT_{(2)}+\frac{1}{3}\partial^2\beta_{(2)}+T\lh \dot{\alpha}+\dot{\tilde{\alpha}}\rh 
+\frac{1}{3}\partial^i\lh T(N_{i}+\tilde{N}_{i})\rh +\beta^k\partial_k\lh \dot{\alpha}+\dot{\tilde{\alpha}}\rh, 
\eea
while acting with $D_{ij}$ we find the form of $\partial^2\beta_{(2)}$:
\bea 
\frac{1}{3}\partial^2\beta_{(2)} & = &
\frac{1}{3}\lh\partial^2\tilde{F}_{(2)}-\partial^2F_{(2)}\rh 
+\frac{1}{6}\partial^iT(N_{i}+\tilde{N}_{i})
-\frac{1}{2}\partial^{-2}\partial^i\partial^j\lh \partial_{(i}T(N_{j)}+\tilde{N}_{j)}) \rh 
\nn\\ 
&&\!\!\!\!-
\frac{1}{2}\partial^{-2}\partial^i\partial^j
\lh \beta^\mu\partial_\mu(\chi_{ij}+\tilde{\chi}_{ij})
\rh
-\frac{1}{4}\partial^{-2}\partial^i\partial^j\lh
\beta^k_{,i}\beta_{k,j}
-\beta_i^{,k}\beta_{j,k}\rh \nn\\ 
&&\!\!\!\!
-\partial^{-2}\partial^i\partial^j\lh
\chi_{k(j}(\beta^k_{,i)}-\beta^{,k}_{i)})
\rh 
. 
\eea 
Combining the above equations we finally find
\be 
\frac{1}{2}\tilde{\alpha}_{(2)}=\frac{1}{2}\alpha_{(2)} 
+\frac{1}{2}NHT_{(2)}
+\frac{1}{2}T\lh \dot{\alpha}+\dot{\zeta}\rh 
+\partial^2\cB,
\ee 
with
\bea
&&\partial^2\cB=\frac{1}{2}\beta^k\partial_k\lh \dot{\alpha}+\dot{\tilde{\alpha}}\rh 
+\frac{1}{6}T\partial^2\tilde{F}
+\frac{1}{4}\Big[\partial^iT(N_{i}+\tilde{N}_{i})
-\partial^{-2}\partial^i\partial^j\lh \partial_{(i}T(N_{j)}+\tilde{N}_{j)}) \rh \Big]
\nn\\
&&\qquad\ \ 
-
\frac{1}{4}\partial^{-2}\partial^i\partial^j
\lh \beta^\mu\partial_\mu(\chi_{ij}+\tilde{\chi}_{ij})
\rh
-\frac{1}{8}\partial^{-2}\partial^i\partial^j\lh
\beta_{k,i}\beta^k_{,j}
-\beta_{i,k}\beta_{j}^{,k}\rh \nn\\ 
&&\qquad\ \ 
-\frac{1}{2}\partial^{-2}\partial^i\partial^j\lh
\chi_{k(j}(\beta_{k,i)}-\beta_{i),k})
\rh
+\frac{1}{6}\lh \partial^2\tilde{F}_{(2)}
-\partial^2F_{(2)}\rh 
. 
\eea
All the terms in $\partial^2\cB$ vanish on super-horizon scales: spatial 
gradients can be neglected in that case, and the divergenceless vectors 
$\beta_i$ and $N_i$ and the transverse traceless tensor $\chi_{ij}$ rapidly 
decay on super-horizon scales (see e.g.~\cite{Rigopoulos:2005xx}).

\subsection{Second-order $0i$ Einstein equation}

In order to simplify calculations we choose to work in a gauge where $F_{(1)}=F_{i(1)}=0$. Consequently we can set $N_{i\perp(1)}=0$ (see \cite{Tzavara:2011hn}). 
The second-order $0i$ Einstein equation takes the form 
\bea
&&\frac{1}{4}\partial^2N_{i\perp(2)}-\frac{1}{N}\partial_kN_{(1)}\partial^k\partial_i\psi_{(1)}
-2\frac{a^2H}{N}\partial_k
\psi_{(1)}\partial_i\partial^k\psi_{(1)} 
-\partial^2\alpha_{(1)}\partial_i\psi_{(1)}
-2\partial_i\partial^k\alpha_{(1)}\partial_k\psi_{(1)}
\nn\\
&&
+\partial_k\alpha_{(1)}\partial_i\partial^k\psi_{(1)}
-\partial_i\alpha_{(1)}\partial^2\psi_{(1)}
-H\partial_iN_{(2)}+\partial_i\dot{\alpha}_{(2)}
+6\frac{H}{N}N_{(1)}\partial_iN_{(1)}
-\frac{2}{N}\dot{\alpha}_{(1)}\partial_iN_{(1)}
\nn\\
&&-\frac{4}{N}N_{(1)}\partial_i\dot{\alpha}_{(1)}
=
\nn\\
&& 
\e NH\partial_iQ_{(2)}-\frac{2\e}{c_{ad}^2}\lh\dot{\zeta}_{1(1)}
-\dot{\alpha}_{(1)}\rh\partial_i\lh\zeta_{1(1)}
-\alpha_{(1)} \rh-\e\lh\frac{\e}{c_{ad}^2}+\hpa\rh NH\partial_i\lh\zeta_{1(1)}
-\alpha_{(1)}\rh^2\nn\\
&&-2\e H N_{(1)}\lh\frac{1}{c_{ad}^2}+1\rh \partial_i\lh\zeta_{1(1)}
-\alpha_{(1)}\rh 
-\e\lh\e+\tilde{\eta}^\parallel\rh NH\partial_i\zeta_{2(1)}^2
-2\e\dot{\zeta}_{2(1)}\partial_i\zeta_{2(1)}\nn\\
&&
-2\e\hpe NH
(\zeta_{1(1)}-\alpha_{(1)})\partial_i\zeta_{2(1)}
+\frac{2\e}{c_{ad}^2}(\hpe-\Lambda^\perp c_{ad}^2) NH\zeta_{2(1)}
\partial_i(\zeta_{1(1)}-\alpha_{(1)}),
\label{app0ieq} 
\eea
where we re-expressed $\varphi_{(1)}$ in terms of the first-order adiabatic and isocurvature  perturbations. The first-order $\psi_{(1)}$ and $N_{(1)}$ have the form (see \cite{Tzavara:2011hn})
\be 
\partial^2\psi_{(1)}=-\frac{N}{Ha^2}\partial^2\alpha_{(1)}+\partial^2\lambda 
\quad\mathrm{and}\quad
N_{(1)}=
\frac{\dot{\alpha}_{(1)}}{H}-\e N(\zeta_{(1)}-\alpha_{(1)}).
\ee 
To eliminate the $\alpha_{(2)}$ and $N_{(2)}$ we 
perturb $NH=\dot{a}/a$ to second order, 
\be 
\frac{\dot{\alpha}_{(2)}}{H}-N_{(2)}=
\frac{N}{H}H_{(2)}+\frac{2}{H}N_{(1)}H_{(1)}.
\ee
Next we rewrite $H_{(2)}$ in terms of $\rho_{(2)}$ taking into account that 
$H_{(1)}=\e H\lh\zeta_{(1)}-\alpha_{(1)}\rh$:
\be
\frac{N}{H}H_{(2)}=-\e N^2H\frac{\rho_{(2)}}{\dot{\rho}}
-\e^2N\lh\zeta_{(1)}-\alpha_{(1)}\rh^2. 
\ee
Inserting these results into (\ref{app0ieq}) we can conclude that
\be
\partial^2N_{i\perp(2)}=0
\ee
and
\bea
&&\!\!\!\!\!\!\!\!\!\!\!\!\!\!\!\!\!\!
-\e NH\Big[
Q_{1(2)}+NH\frac{\rho_{(2)}}{\dot{\rho}}
+(\e-\hpa)\lh\zeta_{1(1)}
-\alpha_{(1)}\rh^2
-(\e+\tilde{\eta}^\parallel)\zeta_{2(1)}^2
-2\hpe\zeta_{2(1)}\lh\zeta_{1(1)}
-\alpha_{(1)}\rh 
\nn\\&&
-\frac{2}{NH}\partial^{-2}\partial_i\dot{\zeta}_{2(1)}
\partial_i\zeta_{2(1)}\Big]
+\partial^2\cA
=0,
\label{r2}
\eea
where 
\bea
&&\!\!\!\!\!\!\!\!\!\!\!\!\!\!\!\!\!\!\partial^2\cA=\partial^{-2}\partial^i\Bigg[
2\partial^2\lambda\partial_i\lh\zeta_{1(1)}
-\alpha_{(1)}\rh 
-\frac{1}{N}\partial^kN_{(1)}\partial_k\partial_i\psi_{(1)}
-2\frac{a^2H}{N}\partial^k
\psi_{(1)}\partial_i\partial_k\psi_{(1)}
\nn\\&&\quad\qquad 
-2\partial_i\partial^k\alpha_{(1)}\partial_k\psi_{(1)}
+\partial^k\alpha_{(1)}\partial_i\partial_k\psi_{(1)}
-\partial_i\alpha_{(1)}\partial^2\psi_{(1)}
\Bigg].
\eea
Note that for an arbitrary gauge the only change in expression (\ref{r2}) will be the appearance of additional spatial gradient terms in $\partial^2\cA$.

\section{Details of the computation for DBI inflation and its third-order action}
\label{dbi_intermediate}

Here we briefly summarize some intermediate steps
to obtain the quantities in DBI inflation considered in subsection~\ref{subsec_dbi} and also give an alternative expression for the third-order DBI action. 

For $\bar{Y}$ and $\bar{\gamma}$, we can show that 
\begin{eqnarray}
\bar{Y}_{\langle AB \rangle} 
&=& \bar{G}_{AB}-2\bar{f}( \bar{X} \bar{G}_{AB} - \bar{X}_{AB}) 
\rightarrow \gamma^2 G_{AB} + 2f X_{AB}\,,\\
\bar{Y}_{;A} &=& - \bar{f}_{;A} (\bar{X}^2 - \bar{S}) \rightarrow 0\,,\\
\bar{\gamma}_{\langle AB \rangle} 
&=&-\frac{\bar{f}}{\bar{\gamma}} \bar{Y}_{\langle AB \rangle} 
\rightarrow -\frac{f}{\gamma}(\gamma^2 G_{AB} + 2f X_{AB})\,,\\
\bar{\gamma}_{;A} 
&=&-\frac{\bar{f}}{\bar{\gamma}} \bar{Y}_{;A} - 
\frac{(1-\bar{\gamma}^2)}{2 f \bar{\gamma}} \bar{f}_{;A} 
\rightarrow - \frac{ (1-\gamma^2)}{2 f \gamma}f_{;A}\,,
\end{eqnarray}
which gives
\begin{eqnarray}
\bar{P}_{\langle AB \rangle}  \rightarrow
\frac{1}{\gamma} 
\left(\gamma^2 G_{AB} + 2 f X_{AB}\right)\,,\quad
\bar{P}_{;A} 
\rightarrow  \frac{(1-\gamma)^2}{2 f^2 \gamma} f_{;A}  -W_{;A}\,.
\end{eqnarray}
We remind the reader that the arrow indicates the limit of taking
the homogeneous background value.

Similarly, to obtain the quadratic terms in the expansion of $P$,
we can show that
\begin{eqnarray}
\bar{Y}_{\langle AB \rangle \langle  CD \rangle} 
&=& -2\bar{f}\left(\bar{G}_{AB} \bar{G}_{CD} - 
\frac12 \bar{G}_{AC} \bar{G}_{BD}  
- \frac12 \bar{G}_{AD} \bar{G}_{BC}\right)\nonumber\\
&\rightarrow& 
-2f\left(G_{AB} G_{CD} - 
\frac12 G_{AC} G_{BD}  
- \frac12 G_{AD} G_{BC}\right)\,,\\
\bar{Y}_{\langle AB \rangle ;C}  &=&
-2\bar{f}_{;C} (\bar{X} \bar{G}_{AB} - \bar{X}_{AB})
\rightarrow -2f_{;C} (X G_{AB} - X_{AB})\,,\\
\bar{Y}_{;AB}  &=&  -\bar{f}_{;AB} (\bar{X}^2-\bar{S}) 
\rightarrow 0\,,
\end{eqnarray}
and
\begin{eqnarray}
\bar{\gamma}_{\langle AB \rangle \langle  CD \rangle} 
&=&-\frac{\bar{f}}{\bar{\gamma}} \left(-\frac{1}{\bar{\gamma}}
\bar{Y}_{\langle AB \rangle} \bar{\gamma}_{\langle  CD \rangle}
+\bar{Y}_{\langle AB \rangle \langle  CD \rangle}\right)\nonumber\\
&\rightarrow &-\frac{f^2}{\gamma}
\Biggl(\frac{1}{\gamma^2}(\gamma^2 G_{AB} + 2f X_{AB})
(\gamma^2 G_{CD} + 2f X_{CD})\nonumber\\
&&\quad\quad\quad\quad\quad
-2\left( G_{AB}  G_{CD} -\frac12G_{AC}  G_{BD} -\frac12G_{AD}  G_{BC}   
\right)\Biggr)\,,\\
\bar{\gamma}_{\langle AB \rangle ; C} &=& 
\left(-\frac{\bar{f}_{;C}}{\bar{\gamma}}+\frac{\bar{f}}{\bar{\gamma}^2}
\bar{\gamma}_{;C}\right) \bar{Y}_{\langle AB \rangle}
-\frac{\bar{f}}{\bar{\gamma}}\bar{Y}_{\langle AB \rangle ;C}\nonumber\\
&\rightarrow &
-\frac{ (1+\gamma^2) f_{;C}}{2 \gamma^3} (\gamma^2 G_{AB} 
+2f X_{AB})
+\frac{2f f_{;C}}{\gamma} (X G_{AB} -X_{AB} )\,,\\ 
\bar{\gamma}_{; AB} &=& 
\left(-\frac{\bar{f}_{;B}}{\bar{\gamma}}+
\frac{\bar{f} \bar{\gamma}_{;B}}{\bar{\gamma}^2}\right) \bar{Y}_{;A}
-\frac{\bar{f}}{\bar{\gamma}} \bar{Y}_{;AB} +
\frac{1+\bar{\gamma}^2}{2 \bar{f} \bar{\gamma}^2} \bar{\gamma}_{;B} 
\bar{f}_{;A}
+\frac{1-\bar{\gamma}^2}{2 \bar{f}^2 \bar{\gamma}} \bar{f}_{;A} \bar{f}_{;B}
-\frac{1-\bar{\gamma}^2}{2 \bar{f} \bar{\gamma}} \bar{f}_{;AB}
\nonumber\\
&\rightarrow&
-\frac{(1-\gamma^2)^2 f_{;A} f_{;B}}{4 f^2 \gamma^3}
-\frac{(1-\gamma^2)f_{;AB } }{2f \gamma}\,.
\end{eqnarray}
It is worth mentioning that during the calculations above,
we must take the background value only at the last step.
If we take this operation before taking derivatives,
we would obtain wrong results. Putting all this together we obtain
\begin{eqnarray}
\bar{P}_{\langle AB \rangle \langle  CD \rangle} 
&\rightarrow &\frac{f}{\gamma}
\biggl(\frac{1}{\gamma^2}(\gamma^2 G_{AB} + 2f X_{AB})
(\gamma^2 G_{CD} + 2f X_{CD})-\nonumber\\
&&\quad\quad\quad\quad
2\left( G_{AB}  G_{CD} -\frac12G_{AC}  G_{BD} -\frac12G_{AD}  G_{BC}   
\right)\biggr)\,,\\
\bar{P}_{\langle AB \rangle  ; C} 
&\rightarrow& \frac{f_{;C}}{\gamma} \left( \frac{(1-\gamma^2)}{2f \gamma^2}
(\gamma^2 G_{AB} +2f X_{AB})- 2(X G_{AB} -X_{AB})
\right)\,,\\
\bar{P}_{ ; AB}
&\rightarrow& \frac{(1-\gamma)^2 f_{;AB} }{2 f^2 \gamma}
+\frac{(1-\gamma)^3 (1+3 \gamma) f_{;A} f_{;B}}
{4 f^3 \gamma^3}-W_{;AB}\,.
\end{eqnarray}

In a similar way, we can obtain the cubic terms in the expansion of $P$ 
as
\begin{eqnarray}
\bar{P}_{\langle AB \rangle \langle  CD \rangle \langle  EF \rangle} 
&\rightarrow & 3 \frac{f^2}{\gamma^5} (\gamma^2 G_{AB} + 2f X_{AB})
(\gamma^2 G_{CD} + 2f X_{CD}) (\gamma^2 G_{EF} + 2f X_{EF})\nonumber\\
&&-2 \frac{f^2}{\gamma^3} (\gamma^2 G_{AB} + 2f X_{AB}) 
(G_{CD} G_{EF} -\frac12 G_{CE} G_{DF}-\frac12 G_{CF} G_{DE})
\nonumber\\
&&-2 \frac{f^2}{\gamma^3} (\gamma^2 G_{CD} + 2f X_{CD}) 
(G_{EF} G_{AB} -\frac12 G_{EA} G_{FB}-\frac12 G_{EB} G_{FA})
\nonumber\\
&&-2 \frac{f^2}{\gamma^3} (\gamma^2 G_{EF} + 2f X_{EF}) 
(G_{AB} G_{CD} -\frac12 G_{AC} G_{BD}-\frac12 G_{AD} G_{BC})\,,\\
\bar{P}_{\langle AB \rangle \langle  CD \rangle ;E} 
&\rightarrow & -\frac{f_{;E}}{\gamma}
\Biggl(\frac{1}{\gamma^2}(\gamma^2 G_{AB} + 2f X_{AB})
(\gamma^2 G_{CD} + 2f X_{CD})\nonumber\\
&&\quad\quad\quad\quad
-2\left( G_{AB}  G_{CD} -\frac12G_{AC}  G_{BD} -\frac12G_{AD}  G_{BC}   
\right)\Biggr)\nonumber\\
&&+\frac{f_{;E}}{2 \gamma^5} (3+\gamma^2) 
(\gamma^2 G_{AB} + 2f X_{AB}) (\gamma^2 G_{CD} + 2f X_{CD})\nonumber\\
&&
-\frac{2f f_{;E}}{\gamma^3} 
\bigl((X G_{AB} -X_{AB})(\gamma^2 G_{CD} + 2f X_{CD})\nonumber\\
&&\quad\quad\quad\quad+(\gamma^2 G_{AB} + 2f X_{AB})(X G_{CD} -X_{CD})\bigr)\nonumber\\
&&-\frac{f_{;E } }{\gamma^3} (1+3 \gamma^2) 
(G_{AB} G_{CD} -\frac12 G_{AC} G_{BD}-\frac12 G_{AD} G_{BC})\,,\\
\bar{P}_{\langle AB \rangle ;CD} 
&\rightarrow& \left(\frac{(1-\gamma^2) f_{;CD}}{2 f \gamma^3}
+\frac{3 (1-\gamma^2)^2 f_{;C} f_{;D}}{4 f^2 \gamma^5}\right)
(\gamma^2 G_{AB} + 2f X_{AB})\nonumber\\
&&+\left(-\frac{2}{\gamma} f_{;CD}
-\frac{2 (1-\gamma^2) f_{;C} f_{;D}}{f \gamma^3}\right)
(X G_{AB} -X_{AB})\,,\\
\bar{P}_{;ABC} 
&\rightarrow&
\frac{(1-\gamma )^2}{2 f^2 \gamma}f_{;ABC}
+\frac{(1-\gamma )^3 (1+3 \gamma )}{4 f^3 \gamma^3}
(f_{;AB} f_{;D}+f_{;BC} f_{;A} +f_{;CA} f_{;B} )\nonumber\\
&&
+\frac{3 (1-\gamma )^4 (1+4 \gamma + 5 \gamma^2)}{8 f^4 \gamma^5}
f_{;A} f_{;B} f_{;C} -W_{;ABC}\,.
\end{eqnarray}
Using the above expressions one can easily recover the DBI quantities given in subsection  \ref{subsec_dbi}.

We also give here an alternative expression for the third-order DBI action. Perturbing the ADM action in the flat gauge to third order,  replacing field perturbations by adiabatic and isocurvature perturbations according to (\ref{gaugezeta}) and using the slow-roll parameter results from section~\ref{subsec_dbi} one can find the explicit exact form
\bea
S_{3(1)}&=& \int d^4 x \frac{a^3\e}{\kappa^2N}\Bigg\{
-\frac{\dot{Z}_1}{NH}
\Bigg[\frac{\dot{Z}_1^2}{c_{s}^2}
\lh \frac{1}{c_s^2}-1\rh  
+\dot{Z}_2^2\lh\frac{1}{c_s^2}-1\rh 
\Bigg] \nn\\
&+&\dot{Z}_1^2
\Bigg[\e\frac{\zeta_1}{c_{s}^2}
\lh \frac{3}{c_s^2}-2\rh-\frac{3}{c_s^2} \Lambda_{\bar{0}}^m
\zeta_m\Bigg]+\dot{Z}_2^2
\Bigg[\frac{\e}{c_s^2} \zeta_1 - \Lambda_{\bar{0}}^m \zeta_m\Bigg]
\nn\\
&-&NH\dot{Z}_1\Bigg[
\frac{2}{3}\e\hat{R}_{1221}\lh\frac{3}{c_{s}^2}+1\rh\zeta_2^2
+3\Lambda_{\bar{0}}^{mn}\zeta_m\zeta_n
+2\e\zeta_1\lh\frac{\e}{2}\frac{\zeta_1}{c_{s}^2}\lh 
\frac{3}{c_s^2} -1\rh 
-\frac{3}{c_s^2}\Lambda_{\bar{0}}^m \zeta_m\rh
\Bigg]\nn\\
\qquad\quad&+&\frac{8}{3} NH \e \hat{R}_{1221}\dot{Z}_2
\zeta_1\zeta_2
\nn\\
&+&(NH)^2\Bigg[-\frac{2\e}{3}\lh\frac{1}{\kappa}\sqrt{\frac{2\e}{\Xi}}\hat{R}_{1221}^{;m}+ 4\hat{R}_{1221}\Lambda_{\bar{0}}^m\rh\zeta^2_2
\zeta_m-3\Lambda^{lmn}\zeta_l\zeta_m\zeta_n\
-\frac{3 \e^2}{c_s^2}\Lambda_{\bar{0}}^m \zeta_m\zeta_1^2
\nn\\
&&\qquad\quad\ \ \ \ +
\e\lh3(\Lambda_{\bar{0}}^{mn}-\Lambda^{mn})\zeta_m
\zeta_n+\frac{2\e}{c_{s}^2}\hat{R}_{1221}\zeta_2^2\rh 
\zeta_1
+\frac{\e^3}{c_s^4} \zeta_1^3 \Bigg]
\nn\\
&-& 2  \Bigg[\frac{\dot{Z}_1 }{c_{s}^2} 
h^{ij} \partial_i \zeta_1 \partial_j \lambda + \dot{Z}_2 
h^{ij} \partial_i \zeta_2 \partial_j \lambda\Bigg]
+
2NH \Bigg[\frac{\e}{c_{s}^2}\zeta_1 
h^{ij} \partial_i \zeta_1 \partial_j \lambda - \Lambda_{\bar{0}} ^m
\zeta_m h^{ij} \partial_i \zeta_1 \partial_j \lambda\Bigg]
\nn\\
&&+\frac{N}{ H} \Bigg[\lh \frac{1}{c_{s}^2}-1 \rh
\dot{Z}_1 h^{ij} \partial_i \zeta_1 \partial_j \zeta_1 
-(1-c_{s}^2)\dot{Z}_1
h^{ij} \partial_i \zeta_2 \partial_j \zeta_2
+2(1-c_{s}^2) \dot{Z}_2 h^{ij} \partial_i \zeta_1 \partial_j \zeta_2\Bigg]
\nn\\
&&+N^2 \Bigg[\e \lh 2-\frac{1}{c_{s}^2} \rh \zeta_1 
h^{ij} \partial_i \zeta_1 \partial_j \zeta_1+ \Lambda_{\bar{0}} ^m
\zeta_m h^{ij} \partial_i \zeta_1 \partial_j \zeta_1
+\e \zeta_1 h^{ij} \partial_i \zeta_2 \partial_j \zeta_2 
 +\Lambda_{\bar{2}} ^m \zeta_m h^{ij} 
\partial_i \zeta_2 \partial_j \zeta_2 
\Bigg]\nn\\
&&+ \Bigg[-3 \e N^2 H^2 \zeta_1^3 
+ 2 \e N H (\partial^2 \lambda) \zeta_1^2 + \frac{1}{2}
(h^{ik} h^{jl} \partial_i \partial_j \lambda \partial_k \partial_l \lambda
- (\partial^2 \lambda)^2)\zeta_1 \Bigg]\Bigg\},
\label{DBI_third_exact}
\eea
where we remind the reader that $\dot{Z}_m$ is given in (\ref{zcap}).

\bibliography{bibsh}{}

\bibliographystyle{utphys.bst}

\end{document}